\documentclass[fleqn,usenatbib]{mnras}

\usepackage{newtxtext,newtxmath}

\usepackage[T1]{fontenc}
\usepackage{ae,aecompl}
\usepackage{appendix,float}
\usepackage{float} 
\usepackage[section]{placeins}
\usepackage{rotating}
\usepackage{threeparttable}
\usepackage{longtable}
\usepackage{xtab,booktabs}
\usepackage{caption}
\usepackage{subcaption}
\usepackage{float}
\usepackage{chemformula}

\usepackage{graphicx}	
\usepackage{amsmath}	

\title[Polycyclic aromatic hydrocarbons and the ionized gas]{Polycyclic aromatic hydrocarbons and the ionized gas in galaxies with active nuclei }
\author[A.~Silva-Ribeiro et al.]{
A. Silva-Ribeiro,$^{1}$\thanks{E-mail:adrianaufma@yahoo.com.br (KTS)}
A.~C.~Krabbe,$^{1}$
C.~M.~Canelo,$^{2}$
A.~F.~Monteiro,$^{1}$
Dinalva~A.~Sales$^{3}$ 
\newauthor J. A. Hernandez-Jimenez$^{1}$ and Andrade,~D.~P.~P.$^{4}$
\\
$^{1}$Universidade do Vale do Para\'{\i}ba, Av. Shishima Hifumi, 2911, Cep 12244-000, S\~ao Jos\'e dos Campos, SP, Brazil\\
$^{2}$Instituto de Astronomia, Geofísica e Ciências Atmosféricas, Departamento de Astronomia, Universidade de São Paulo, 05508-090, Brazil.\\
$^{3}$Instituto de Matemática, Estatística e Física, Universidade Federal do Rio Grande, 96203-900, RS, Brazil. \\
$^{4}$Observatório do Valongo, Universidade Federal do Rio de Janeiro, Rio de Janeiro, 20080-090, Brazil.\\
}

\date{Accepted XXX. Received YYY; in original form ZZZ}

\pubyear{2020}

\begin{document}
\label{firstpage}
\pagerange{\pageref{firstpage}--\pageref{lastpage}}
\maketitle

\begin{abstract}
We present a study for a sample of galaxies with active nuclei to characterize the main type of PAH molecules present in these objects and the local physical conditions of their irradiating sources, as well as the characteristics of the residing ionized gas, by combining optical and infrared data. Photoionization models were built with the  {\sc CLOUDY} code to reproduce optical emission line ratios in combination with PAH intensity ratios. We find that the species containing 10 $-$ 82 carbon atoms are the most abundant in the sample. We suggest  that family of species with only two or three fused rings of and a nitrogen hanging, such as small aromatic amides are important targets worthy of consideration in future experimental/theoretical as well as observational studies. We find that the AGN photoionization models reproduce most of the observational data in the log (6.2/11.3) versus log ([\ion{N}{ii}]$\lambda$6584/H$\alpha$) diagram with  the optical to X-ray spectral index of $\alpha_{\rm ox}=-1.4$. The flux of small PAH, as well as the flux of ionized PAHs and PANH, decrease as the logarithm of the ionization parameter (log $U$) increases.  The 6.2/11.3 PAH intensity ratio presents anti correlation between the oxygen abundance and log $U$, in the sense that the 6.2/11.3 ratio decreases as the oxygen abundance and log $U$ increases. Finally, we found  that the ionization degree of PAH species increases with the decreasing of the 11.3/7.7
ratio and the  log U, in agreement with the models proposed by Draine \& Li .

\end{abstract}




\section{Introduction}

Polycyclic Aromatic Hydrocarbons (PAHs) are organic compounds that contain two or more fused benzene rings and can also form large aromatic structures. 
Although the term PAHs refers to molecules containing only carbon and hydrogen atoms, the more general term "polycyclic aromatic compounds" also includes the functional derivatives, just like nitro-PAHs, Oxy-PAHs, and the heterocyclic analogues. Studies show that PAHs represent one of the most abundant forms of carbon in the Universe, with approximately 15 per cent of the elemental carbon of the Universe is in the form of PAHs \citep{candian2019aromatic}. Besides,  PAHs have an important role in heating the Interstellar Medium (ISM) since they have low ionization potential  \citep{tielens2008interstellar}. This class of molecules is often observed in a wide range of astrophysical environments, including extragalactic sources such as Seyferts and Starbursts galaxies \citep{brandl2006mid,smith2007mid,sales2010polycyclic,sales2013polycyclic}. They present stronger emission bands at 3.3, 6.2, 7.7, 8.6, 11.2,  and 12.7~$\mu$m wavelength \citep[e.g.][]{li2020spitzer}.

There is an astrobiological interest in PAHs due to their potential role in the formation of protocells, which may help to evaluate possible pathways for the origins of life on the early Earth. Like PAHs, Nitrogen-containing polycyclic aromatic hydrocarbons, known as PANHs, can fragment and form prebiotic molecules, such as ethanol and amino acids \citep{ehrenfreund2006experimentally}. In addition, such species can self-assemble, originating from the formation of protocells \citep{schrum2010origins}. 

\citet{canelo2018variations} analyzed the 6.2~$\mu$m PAH emission band for 155 predominantly starburst-dominated galaxies and distributed it into the Peeters’ A, B and C classes \citep{peeters2002rich}, revealing a predominance of 67 per cent of class A sources with band profiles corresponding to a central wavelength near 6.22$\mu$m, which can only be explained by PANHs \citep{hudgins2005variations}. In this way, the PANHs, in addition to nitrogen in the gas phase and ices of the ISM, could be another reservoir of nitrogen in the Universe.

Studies have shown that the physical conditions of the object under study at radiation and metallicity levels, for example, affect both approaches. \citet{hanine2020formation} found that the size of the formed PAH species increases roughly with increasing temperature up to 800~K,  forming rings up to 32 carbon atoms and being correlated with the level of dehydrogenation, which can make these molecules highly reactive \citep{chen2018formation},  characteristic that can intensify the formation of large PAHs.1 \citet{seok2014formation}, for example, concluded that the PAH abundance is low at low metallicities (Z~<~0.1~$Z{_\odot}$), but increases rapidly beyond a certain metallicity (Z~>~0.3~$Z{_\odot}$) by chemical evolution model in galaxies. 
On other hand, \citet{maragkoudakis2018pahs} found that variations on the 17~$\mu$m PAH band depend on the object type, however, there is no dependence on metallicity for both extragalactic  \ion{H}{ii} regions and galaxies.

PAHs are also important molecules from the galactic diagnosis perspective because their emissions can be used as a calibrator for star formation in galaxies (e.g. \cite{shipley2016new,maragkoudakis2018pahs}). Such peculiarity has been a motivating factor for studying the presence of PAHs in active galaxy nuclei (AGNs). Seyfert galaxies are a type of AGNs whose notoriety show presence of PAH emission lines. The ionization of PAH molecules, as well as ionized gas surrounding central engine of the Seyfert galaxies, is dominated by the hardness radiation coming from accreting disk of the supermassive black hole environment. Interestingly, \citet{madden2006ism} and \citet{maragkoudakis2018pahs} have found that the total  PAH/Very Small Grains (VSG) intensity ratio decreases with increasing radiation hardness, as traced by  [\ion{Ne}{iii}]/[\ion{Ne}{ii}],  indicating  that harder  radiation  fields are largely responsible for the destruction of PAHs in low-metallicity regions. In addition, \citet{sales2010polycyclic,sales2013polycyclic} have studied 186 active and star-forming galaxies and could infer that active galaxies seems to present larger PAH molecules ($\geq$ 180 carbon atoms) than starforming galaxies. They propose that PAH molecules can survive near to AGNs since their dusty tori is likely to provide column densities necessary to shield PAHs from the hard radiation field \citep[see also][]{Alonso2014,Ruschel2014}.

The nuclear region of the Seyfert and Starburst are excellent objects for studying the signatures of PAHs in order to map the main species of PAHs that are contributing to the emissions in the IR of Seyfert and Starburst galaxies. In additionally, the presence of metallic emission lines, such as  [\ion{O}{II}] $\lambda$~3727, [\ion{O}{III}] $\lambda$~4363, [\ion{O}{III}] $\lambda$~4959, [\ion{O}{III}] $\lambda$~5007, [\ion{N}{II}] $\lambda$~5755, [\ion{N}{II}] $\lambda$~6548, [\ion{N}{II}] $\lambda$~6584, [\ion{S}{II}] $\lambda$~6717 and [\ion{S}{II}] $\lambda$~6731, which are present in Seyfert and Starburst galaxies make it possible to estimate the physical conditions  (oxygen abundance, density and electronic temperature) of the ionized gas regions. Besides, the ratio of emission lines in combination with photoionization models can be used to determine the hardness of ionizing source and the metallicity of the gas. In this context, it is the possible to verify correlations between the ionized gas regions and the photodissociation regions where PAHs persist.

Concerning the emission features at IR and optical wavelengths, studies have shown (e.g \citet{schlemmer2014laboratory}) that large carbon-based molecules (with $\sim$ 10~--~100 atoms) in the gas phase, such as PAHs, long carbon chains or fullerenes should be possible carriers of Diffuse Interstellar Bands (DIBs) and the Unidentified Infrared (UIR) bands. However, the identification of the carriers of DIBs remains to be examined, with the exception of five bands attributed to C$_{60}^{+}$ \citet{tielens2013molecular,campbell2015laboratory}.  In this sense, the analysis of potential emitting molecules of the PAH population in astrophysical sources in comparison with optical data observed, for instance, can contribute to improving our understanding of those carries and the PAH chemistry in the Universe. 

In this paper, we combined optical and infrared data of a sample of Seyfert (Sy) and Starburst (SB) galaxies to characterize the main type of PAH molecules present
in these objects and the local physical conditions of their irradiating sources, as well as the characteristics of the residing ionised gas. The paper is organized as follows. In Section \ref{sec:data} we describe the selection of our sample and the data utilized. 
In the Section \ref{sec:data_analisis} and \ref{subsection: analysis_optical} the data analysis is presented. The results and discussions are given in Section \ref{sec:results_discussion}.  The
conclusion of the outcome is presented in Section \ref{sec:conclusions}.

\section{Observational Data and Sample Description}
\label{sec:data}

We compiled from the literature a sample of Seyfert and Starburst galaxies with observed and reduced spectra in the infrared and optical.  We selected  all Starburst and Seyfert galaxies with redshift lower than 0.04 from the Spitzer/IRS ATLAS project\footnote{\url{http://www.denebola.org/atlas/}} \citep{hernan2011atlas} and with low-resolution (R $\sim$ 100) spectra in the range from 5 to 15~$\mu$m,  obtained with the Infrared Spectrograph \citep[IRS, ][]{houck2004infrared} instrument on board the Sptizer Space Telescope \citep{Werner04}. From this preliminary sample, we chose only objects with optical spectra in the Sloan Digital Sky Survey (SDSS) DR16 \citep{ahumada2019sixteenth}. Also, all the selected spectra approximately correspond to nuclear region of the galaxies. All the infrared spectra were obtained in staring-mode, except for \citet{buchanan2006spitzer} and \citet{wu2009spitzer}, who utilized the spectral mapping mode, but only the nuclear spectrum was given.  We used short-low spectroscopy module covering interval between 5.2$\mu$m and 14.5$\mu$m with a resolving power of R $\sim$ 60-127 and slit size of 3.6 to 3.7 arcseconds per 57 arcseconds.

\begin{table*}
\caption{Selected objects and their respective information, including the source's name, IR reference, nuclear type,  equatorial coordinates ($\alpha$ and $\delta$), absolute  magnitude in visual band ($M_{\textrm B}$), bolometric luminosity in visual band ($L_{\textrm V}$) and  infrared luminosity $L_{\textrm{IR}}$.}
\label{tab:propriedades_gerais_amostra} 
\begin{tabular}{lrrrcccccc}
\noalign{\smallskip}
\hline
\hline
\noalign{\smallskip}
\multicolumn{1}{c}{Object} & IR Reference & \multicolumn{1}{c}{Nuclear Type} &  
$\alpha(2000)^{a}$ &  $\delta(2000)^{a}$ & $z^{a}$ & \multicolumn{1}{c}{$M_{\textrm B}$(mag)$^{a}$} & \multicolumn{1}{c}{$L_{\textrm V}$ ($L_{\odot}$)$^{a}$}  & \multicolumn{1}{c}{$L_{\textrm {IR}}$ ($L_{\odot}$)$^{a}$}\\
\noalign{\smallskip}
\hline
\noalign{\smallskip}
  Mrk\,273 & \citet{wu2009spitzer} & Seyfert ${[1]}$         & $13^{\rm{h}}44^{\rm{m}}42\fs1$  & $55^{\rm{h}}53^{\rm{m}}13 \fs$0 & 0.038 & $-22.34$ & 8.37$\times 10^{10}$ & 7.94$\times 10^{10}$\\
 MrK\,471 & \citet{deo2007spitzer} & Seyfert ${[1]}$         & $14^{\rm{h}}22^{\rm{m}}55\fs4$ & $32^{\rm{h}}51^{\rm{m}}03\fs0$ & 0.034 & -22.80 & 4.58$\times 10^{10}$ & 4.21$\times 10^{10}$\\
  Mrk\,609 & \citet{deo2007spitzer} & Seyfert ${[1]}$           & $03^{\rm{h}}25^{\rm{m}}25\fs3$ & $-06^{\rm{h}} 08^{\rm{m}} 38\fs$0 & 0.034 & -22.48 & 6.58$\times 10^{10}$ & 4.85$\times 10^{10}$\\
Mrk\,622 & \citet{deo2007spitzer} & Seyfert ${[1]}$           & $08^{\rm{h}}07^{\rm{m}}41\fs0$ & $39^{\rm{h}}00^{\rm{m}}15\fs0$ & 0.029 & -21.44 & 3.18$\times 10^{10}$ & 1.85$\times 10^{10}$ \\
Mrk\,883 & \citet{deo2007spitzer} & Seyfert ${[1]}$            & $16^{\rm{h}}29^{\rm{m}}52\fs9$ & $24^{\rm{h}}26^{\rm{m}}38\fs0$ & 0.038 & -22.19 & 2.70$\times 10^{10}$ & 3.95$\times 10^{10}$\\
NGC\,660 & \citet{brandl2006mid} & LINER ${[1]}$          & $01^{\rm{h}}43^{\rm{m}}02\fs4$  & $13^{\rm{h}}38^{\rm{m}}44\fs4$ & 0.003 & -21.50 & 1.32$\times 10^{10}$ & 2.02$\times 10^{10}$\\
NGC\,2622 & \citet{deo2007spitzer} & Seyfert ${[5]}$            & $08^{\rm{h}}38^{\rm{m}}10\fs9$ & $24^{\rm{h}}53^{\rm{m}}43\fs0$ & 0.023 & -22.80 & 4.75$\times 10^{10}$ & 6.35$\times 10^{10}$ \\
NGC\,2623 & \citet{brandl2006mid} & Starburst ${[2]}$          & $08^{\rm{h}}38^{\rm{m}}24\fs1$  & $25^{\rm{h}}45^{\rm{m}}16\fs9$ & 0.018 & -21.80 & 3.35$\times 10^{10}$ & 1.74$\times 10^{11}$\\
NGC\,4676 & \citet{brandl2006mid} & Starburst  ${[3]}$          & $12 ^{\rm{h}}46^{\rm{m}}10\fs1$  & $30^{\rm{h}}43^{\rm{m}}55\fs0$ & 0.009 & -20.97 & 2.20$\times 10^{10}$ & 5.15$\times 10^{10}$\\
NGC\,4922 & \citet{wu2009spitzer} & Seyfert ${[4]}$        & $13^{\rm{h}}01^{\rm{m}}24\fs9$  & $29^{\rm{h}}18 ^{\rm{m}}40\fs0$ & 0.024 & -22.20 & 2.75$\times 10^{10}$ & ...\\
NGC\,5256 & \citet{wu2009spitzer} & Seyfert ${[4]}$         & $13 ^{\rm{h}}38^{\rm{m}}17\fs5$  & $48^{\rm{h}}16 ^{\rm{m}}37\fs0$ & 0.028 & -22.85 & 5.23$\times 10^{10}$ & 1.06$\times 10^{11}$ \\
 NGC\,5347 & \citet{buchanan2006spitzer} & Seyfert ${[1]}$            & $13^{\rm{h}}53^{\rm{m}}17\fs8$  & $33^{\rm{h}}29^{\rm{m}}27\fs0$ & 0.008& -20.33 &4.65$\times 10^{9}$ & 3.43$\times 10^{9}$\\
\hline
\end{tabular}
\begin{minipage}{17cm}
{\it References}: [1] \citet{veron2003vizier}; [2] \citet{keel1984optical}; [3] \citet{liu1995spectrophotometric}  [4] \citet{rush1993extended}; [5] \citet{deo2007spitzer}.\\
{\it Note}: $^a$ Taken from the NASA Extragalactic Database (NED). 
 \end{minipage}
\end{table*}

\begin{figure}
\includegraphics[width=1\linewidth]{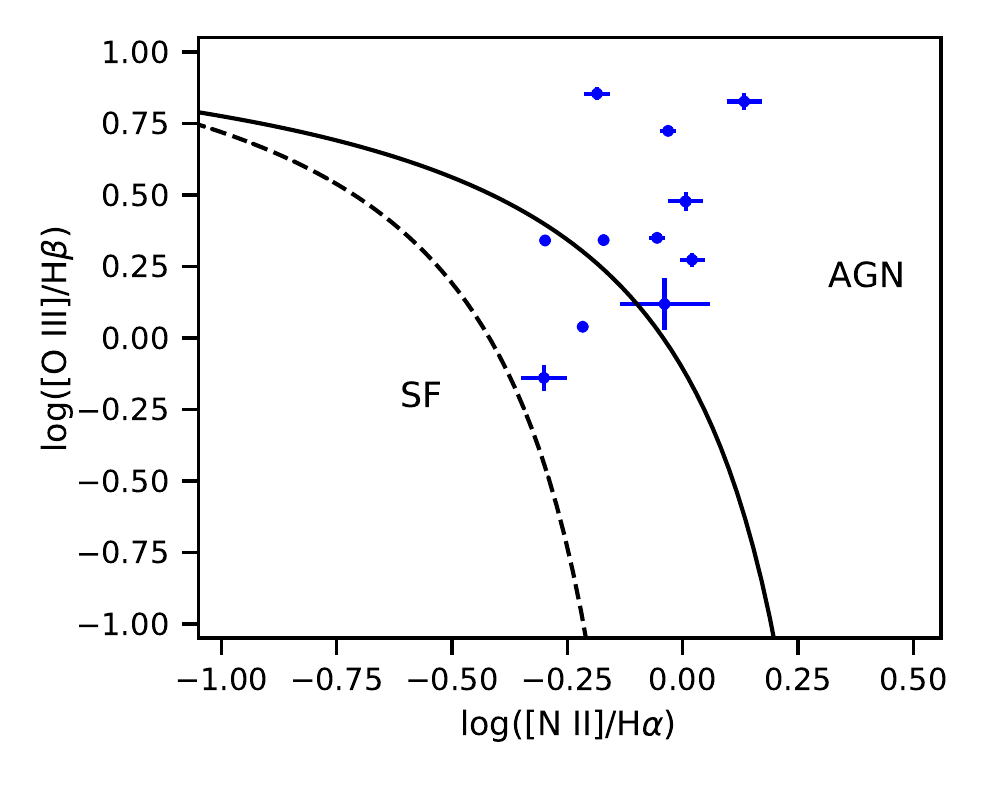}\\
\caption{$[\ion{O}{iii}]/\rm H\alpha$ versus $[\ion{N}{ii}]/\rm H\alpha$ diagnostic diagram for galaxies of our sample. The black solid curve represents the theoretical upper limit for the star-forming regions proposed by \citet{kewley2001theoretical}, and the black dashed curve is the  empirical star-forming limit proposed by \citet{kauffmann2003host}. The region between the Ke01 and Ka03 lines is denominated composite region.}
\label{diagrama:diagnostico}
\end{figure}

Table \ref{tab:propriedades_gerais_amostra} lists the objects selected for this study, including some basic features.
The ionizing source of each galaxy was initially taken from the literature. Diagnostic diagrams 
initially proposed by \citet{baldwin1981classification} (e.g.,  $[\ion{O}{III}]/\rm H\beta$ versus $[\ion{N}{II}]/\rm H\alpha$, 
$[\ion{O}{III}]/\rm H\beta$ versus $[\ion{S}{II}]/\rm H\alpha$, and
$[\ion{O}{III}]/\rm H\beta$ versus $[\ion{O}{I}]/\rm H\alpha$), which are commonly known as BPT diagrams are used  to distinguish objects  ionized by massive stars (SFs), AGN, and Low-ionization nuclear emission-line regions (LINERs). In order to verify the classification of the ionizing source of our sample, we show
in Fig.~\ref{diagrama:diagnostico} the diagram $[\ion{O}{III}]/\rm H\beta$ versus $[\ion{N}{II}]/\rm H\alpha$. The black solid curve represents the theoretical upper limit for the star-forming regions proposed by \citet{kewley2001theoretical}  and the black dashed curve is the empirical star-forming limit proposed by \citet{kauffmann2003host}, hereafter Ke01 and Ka03, respectively. The blue dot-dashed curve represents the  Seyfert-LINER dividing line (Ke01).  
According to Fig.~\ref{diagrama:diagnostico}, some galaxies presented a different classification from the literature. However, the nuclear region of the galaxies of our sample is ionized by AGN, including the galaxies that are between Ke01 and Ka03 lines,  region denominated as composite region, where the object is ionized by SF and AGN.
Then, throughout the text, we will refer to our sample of galaxies as AGNs.

 The optical data compiled from SDSS were obtained with BOSS Spectrograph attached to 2.5 m telescope at Apache Point Observatory. The spectra range considered is between 3\,600 and 10\,400 \AA\,, spectral resolution of  $R =$ 1\,560 at 3\,700  \AA\,, $R =$2\,270 at 6\,000 \AA\, and the fiber diameter is 3 arcsec.

\begin{figure*}
\centering
\includegraphics[width=\textwidth]{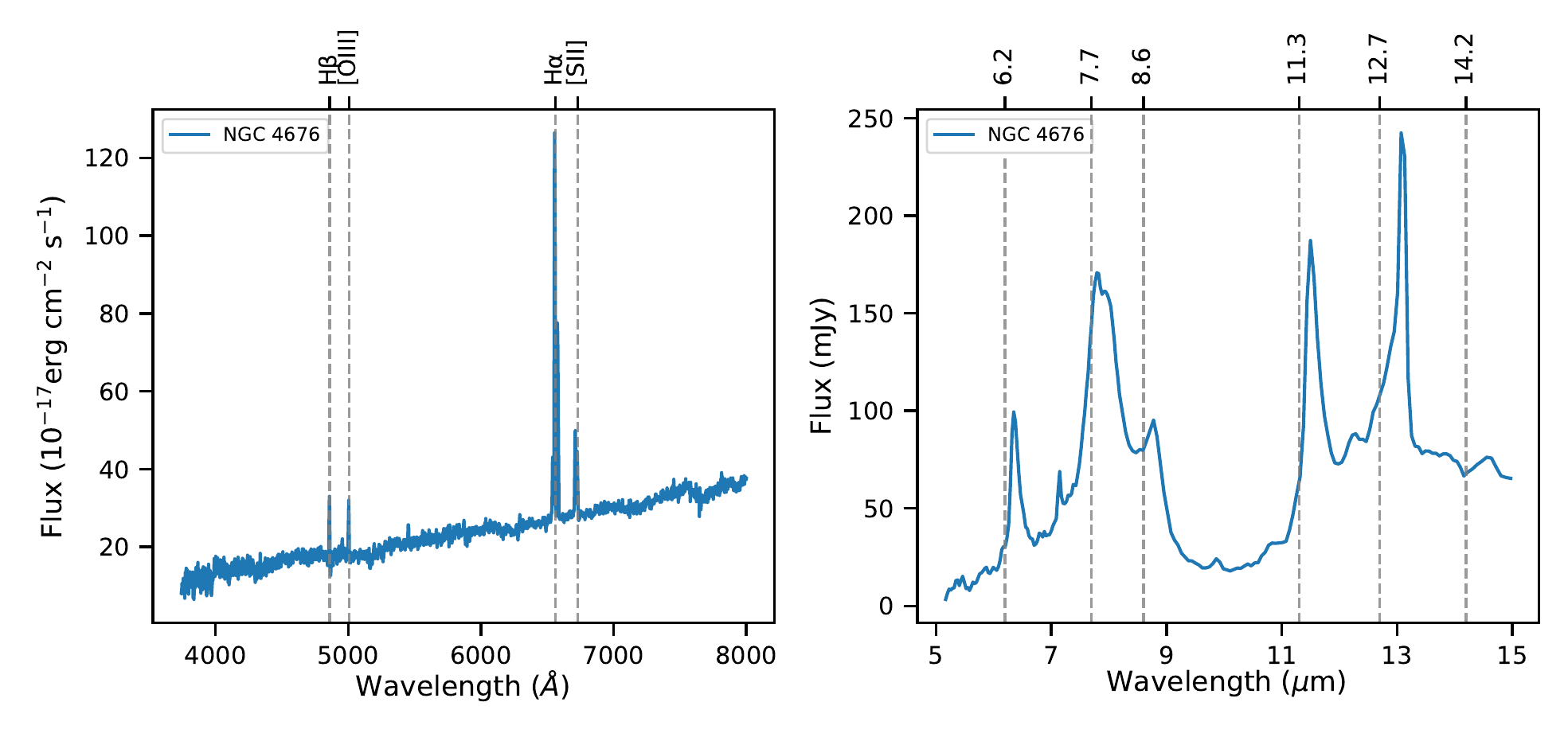}
\caption{Left panel: the observed optical spectrum for the nuclear region of NGC\,4676 obtained with SDSS. Right panel: the observed infrared spectrum for the nuclear region of NGC\,4676 obtained with IRS of Spitzer.}
\label{fig:ngc4676_otico_bruto}
\end{figure*}

Fig.~\ref{fig:ngc4676_otico_bruto} shows an optical spectrum of the nuclear region of the Starburst galaxy NGC\,4676, where some emission lines are highlighted, and an infrared spectrum for the same galaxy with the main bands of PAHs indicated.  All the compiled spectra were reduced and calibrated in flux and wavelength.



\section{Optical data analysis}
\label{subsection: analysis_optical}
The optical spectra from SDSS within the wavelength range of 3\,600-10\,400 \AA\,  were used to estimate some physical properties of the ionized gas through the use of emission line intensity ratios in combination with photoionization models. The spectrum of each galaxy was processed following the steps listed below:

\begin{enumerate}
    
    \item  The spectra were corrected to their rest-frame, adopting the values of $z$ listed in Table \ref{tab:propriedades_gerais_amostra}.
    
    \item The pure nebular  spectra were obtained  after the subtraction of stellar population contribution of the observed spectrum. To obtain the stellar population contribution we use the stellar population synthesis code {\scriptsize\,STARLIGHT} developed by \citet{cid05}; \citet{mateus2006semi}; \citet{asari2007history}. 
    The code fits an observed spectrum with a combination of simple stellar population (SSP) models excluding the emission lines and spurious features (bad pixels or sky residuals). We use the  spectral basis of  \citet{bruzual2003stellar} with 45 synthetic SSPs spectra with three metallicities, Z = [0.2, 1, and 2.5] Z$_\odot $, and 15 ages, t = [0.01, 0.003, 0.005, 0.01, 0.025, 0.04, 0.1, 0.3, 0.6, 0.9, 1.4, 2.5, 5, 11, and 13]$\times 10^{9} $ years. The synthetic SSP spectra have a spectral resolution of 3 \AA\, and to obtain the same spectral resolution of observed spectra these were blended with an elliptical Gaussian function. For more details about the synthesis method see \citet{cid05}, \citet{mateus2006semi}, and \citet{asari2007history}.The flux of the lines in the pure nebular spectra was measuring by using in-house fitting code, which fitting gaussian functions to the line profiles \citep{hernandez13,hernandez15}.

   \item The residual extinction associated with the gaseous component for each galaxy was calculated by comparing the observational  value for H$\alpha$/H$\beta$ ratio to the theoretical value of 2.86 obtained by \citet{hummer1987recombination} 
   for the Case B at  electron temperature and density 
   of 10\,000 K and 100 cm$^{-3}$, respectively.  For this, the following expression was considered
   
   \begin{equation}
   \frac{I(\lambda)}{I(\rm H\beta)}=\frac{F(\lambda)}{\rm F (H\beta)} \times 10^{c(\rm H\beta)[f(\lambda)-f(\rm H\beta)]}
   \end{equation}
   where $I(\lambda)$ is the intensity (reddening corrected) of the emission line at a given wavelength $\lambda$, $F(\lambda)$ is the observed flux of the emission line, $f(\lambda)$ is the adopted reddening curve normalized to $\rm H\beta$, and $c(\rm H\beta$) is the interstellar extinction coefficient.

    \end{enumerate}

Table \ref{tab:fluxo_otico} lists 
the reddening function  $f(\lambda)$,  the reddening corrected emission line intensities $I(\lambda)$, and the logarithmic extinction coefficient $c(\rm H\beta)$ for each object.

\begin{table*}
\caption{Reddening corrected emission-line intensities  and the logarithmic extinction coefficient, $c$(H$\beta$). The  flux for H$\beta$ and H$\alpha$ are 100 and 286, respectively.}
\label{tab:fluxo_otico}
\begin{tabular}{llllllll}
\toprule
Object & [\ion{O}{III}]$\lambda$4959 & [\ion{O}{III}]$\lambda$5007  & [\ion{N}{II}]$\lambda$6548 & [\ion{N}{II}]$\lambda$6584 & [\ion{S}{II}]$\lambda$6716 & [\ion{S}{II}]$\lambda$6731 &     $c$(H$\beta$) \\
$f(\lambda)$ & -0.02 & -0.04 &-0.35  &-0.39 &-0.39 &-0.43 &\\
 \midrule
MrK\,273  &      77$\pm$2&   224$\pm$5  &      87$\pm$2 &   252$\pm$7 &     79$\pm$2 &	   65$\pm$2 &  1.29$\pm$0.02 \\
Mrk\,471  &    226$\pm$10&   670$\pm$28 &     131$\pm$9 &  389$\pm$24 &     96$\pm$8 &	   84$\pm$7 &  0.69$\pm$0.05 \\
Mrk\,609  &    178$\pm$4 &   529$\pm$11 &      90$\pm$3 &   266$\pm$8 &     49$\pm$2 &	   52$\pm$3 &  0.67$\pm$0.03 \\
Mrk\,622  &    101$\pm$9 &   300$\pm$15 &      98$\pm$6 &  290$\pm$18 &     65$\pm$4 &	   56$\pm$4 &  0.67$\pm$0.05 \\
Mrk\,883  &     73$\pm$1 &    219$\pm$1 &      48$\pm$1 &   144$\pm$2 &     74$\pm$1 &	   58$\pm$1 &  0.39$\pm$0.01 \\
NGC\,660  &     66$\pm$4 &    187$\pm$6 &     105$\pm$5 &  300$\pm$13 &     75$\pm$3 &	   58$\pm$3 &  1.55$\pm$0.04 \\
NGC\,2622 &    310$\pm$8 &   947$\pm$23 &     186$\pm$9 &  569$\pm$23 &   175$\pm$10 &	 184$\pm$11 &  0.04$\pm$0.04 \\
NGC\,2623 &     46$\pm$6 &   131$\pm$18 &     91$\pm$14 &  261$\pm$41 &    75$\pm$12 &	   58$\pm$9 &  1.30$\pm$0.14 \\
NGC\,4676 &     25$\pm$3 &     72$\pm$5 &      50$\pm$4 &  143$\pm$12 &     63$\pm$5 &	   44$\pm$4 &  1.29$\pm$0.07 \\
NGC\,4922 &     75$\pm$1 &    220$\pm$2 &      66$\pm$1 &   193$\pm$3 &     50$\pm$1 &	   48$\pm$1 &  1.00$\pm$0.01 \\
NGC\,5256 &     37$\pm$1 &    109$\pm$1 &      58$\pm$1 &   174$\pm$1 &     63$\pm$1 &	   58$\pm$1 &  0.53$\pm$0.01 \\
NGC\,5347 &    239$\pm$8 &   714$\pm$23 &      62$\pm$3 &   186$\pm$9 &     75$\pm$4 &	   66$\pm$4 &  0.55$\pm$0.04 \\

\bottomrule
\end{tabular}

\end{table*}

\subsection{Oxygen abundance and ionization parameter}
 \label{subsection:Oxygen Abundance} 
This section briefly discusses the methods used to estimate the oxygen abundance and the ionization parameter of the ionized gas in the galaxies.

Oxygen is the element widely used as a proxy for global gas-phase metallicity $Z$ \citep{kennicutt2003composition}; \citep{hagele2008precision} of gaseous nebulae. This element has prominent emission lines, and their most important ionization stages are present in the optical spectra of these objects. 
 One of the  most reliable  methods  to determine the oxygen abundance is based on the ratio between oxygen forbidden lines and hydrogen lines, and the electron temperature  ($T_{\rm e}$), specially for star-forming regions and planetary nebulae. Measurements of the auroral lines, such as [\ion{O}{III}]\,$\lambda\,4363$, and [\ion{N}{II}]\,$\lambda\,5755$, are necessary to determine $T_{\rm e}$. Unfortunately, they are very faint and often drop below the detectability level in the spectra of high-metallicity \ion{H}{II} regions. This method is known as the $T_{\rm e}$-method or direct method.  Unfortunately, we do not detect temperature-sensitive emission lines [\ion{O}{III}]\,$\lambda\,4363$, and [\ion{N}{II}]\,$\lambda\,5755$
in the spectra of the galaxies, and then the direct method could no be used to determine the O/H abundance.

Here, we determine the O/H abundance combining the observed emission line ratios  with a grid of photoionization models. The photoionization model grids were built using version 17.02 of the {\sc CLOUDY} code \citep{ferland20172017} in order to compare the observational emission-line intensity ratios for the galaxies in our sample with the photoionization model predictions for similar emission-line intensity ratios. The input parameters of the models are briefly described in the following, however, for more details to these parameters see \citet{dors2015central,dors2017new,dors2020chemical} and \citep{carvalho2020chemical}.

\begin{enumerate}

\item The Spectral Energy Distribution (SED): The SED utilized was typical of AGNs, consisting of the sum from the Big Bump component peaking at $\approx$ 1 Ryd, which is parametrized by the temperature of the bump assumed to be $1.5\:\times\: 10^{5}$ K and, an X-ray power law with spectral index  $\alpha_{\rm x}$ = -1.0,
representing the non-thermal X-ray radiation.  
The continuum between
2 keV and 2500 \AA\, is described by a power law with the optical to X-ray spectral index $\alpha_{ox}$, which  is defined by 
\begin{equation}
      \label{apxeq}
\alpha_{ox}= \frac{\log [F(2\: {\rm keV})/F(2500\: \textrm{\AA})]}{\log [\nu(2 \: {\rm keV})/\nu(2500 \: \textrm{\AA})]}, 
\end{equation}
where $F$ is the flux at 2\:keV,  2500\:\AA\ and $\nu$ are the corresponding frequencies \citep{1979ApJ...234L...9T}.
Therefore,  the $\alpha_{\rm ox}$ is the power-law slope connecting the monochromatic flux at 2500 \AA\, and at 2 keV  and display hard X-ray spectra. For instance, SEDs with small values of 
$\alpha_{\rm ox}$, e.g -2,  represent an ionization source with a very soft spectrum  yielding models with a very low ionization degree   (e.g. \cite{2019MNRAS.486.5853D}).


\item Metallicity ($Z$): The metallicities used in relation with the solar value ($Z/Z_\odot$) was 0.5,  1.0, 1.5, 2.0, 2.5, 3.0, 3.5 and 4.0. The assumed solar oxygen abundance is 12 + log(O/H)$_\odot$ = 8.69 \citep{asplund2009chemical,prieto2001forbidden} whose abundances 8.40, 8.69, 8.86, 8.99,  9.08, 9.16, 9.23, and 9.29, correspond to the above listed metallicities, respectively.

\item Ionization Parameter ($U$): The ionization parameter is defined by {$U$}=${Q}{(\mathrm{H})}/4\pi R^2_0nc$, where the ${Q}{(\mathrm{H}})$ is the number of ionizing photons emitted per second by the ionizing source, $R_0$ is the distance from the ionization source to the inner surface of the ionized gas cloud in cm,  $n$ [cm$^{-3}$]  is the total hydrogen density (ionized, neutral, and molecular), and $c$ is the speed of light in km s$^{-1}$. We considered the logarithm of $U$ in the range of -4.0 $\leq$ log$U$ $\leq$ -1, about the same values considered by \citet{feltre2016nuclear} for AGNs. 

\item Electron Density: We assumed for the models an electron density value of $N_{\rm e}$= 500 cm$^{-3}$, constant in the nebular radius. The electron density ($N_{\rm e}$)  derived from the [\ion{S}{II}]$\lambda\,6717/\lambda\,6731$  for our sample of galaxies is  in the range of 111 to 922 cm$^{-3}$, with a mean density of $N_{\rm e}$= 411 cm$^{-3}$.

\item Inner and outer radius: The inner radius
of 3 pc was assumed, which is the distance from the ionizing source to the illuminated gas region. It is a typical value for Seyfert galaxies \citep{balmaverde2016hst}. The outer radius was assumed to be the one where the electron temperature of the gas reaches 150 K.  It is worth to mentioning that studies has found evidences  that the inner wall of the dust torus has around several tenths of a parsec from the central engine (e.g. \citep{xie2017silicate}). However, models with different combinations of ${Q}{(\mathrm{H}})$, inner radius and $N_{\rm e}$ but that result in the same U are homologous models with about the same predicted emission-line intensities \citep{bresolin1999ionizing} and \citep{2019MNRAS.486.5853D}.


\item PAH: The size distribution for PAHs is given by a power law of the form a$^{-3.5}$, where a is the PAH radius and the number of carbon atoms range from 30 to 500, and  10 size bins \citep{abel2008sensitivity}.

\end{enumerate}

In order to extend this analysis to other sources with a wider range of metallicities, we included data taken from the sample of the Spitzer Infrared Nearby Galaxies Survey (SINGs) 
by \citet{smith2007mid}, in which the 6.2/11.3 and 6.2/7.7 emission ratios were considered.  The log([\ion{N}{ii}]$\lambda$6584/H$\alpha$)  values  of this sample were taken from \citet{2010ApJS..190..233M}.
This sample is composed of Seyfert, LINERS and, star forming galaxies.

\begin{figure}
\includegraphics[width=1.\linewidth]{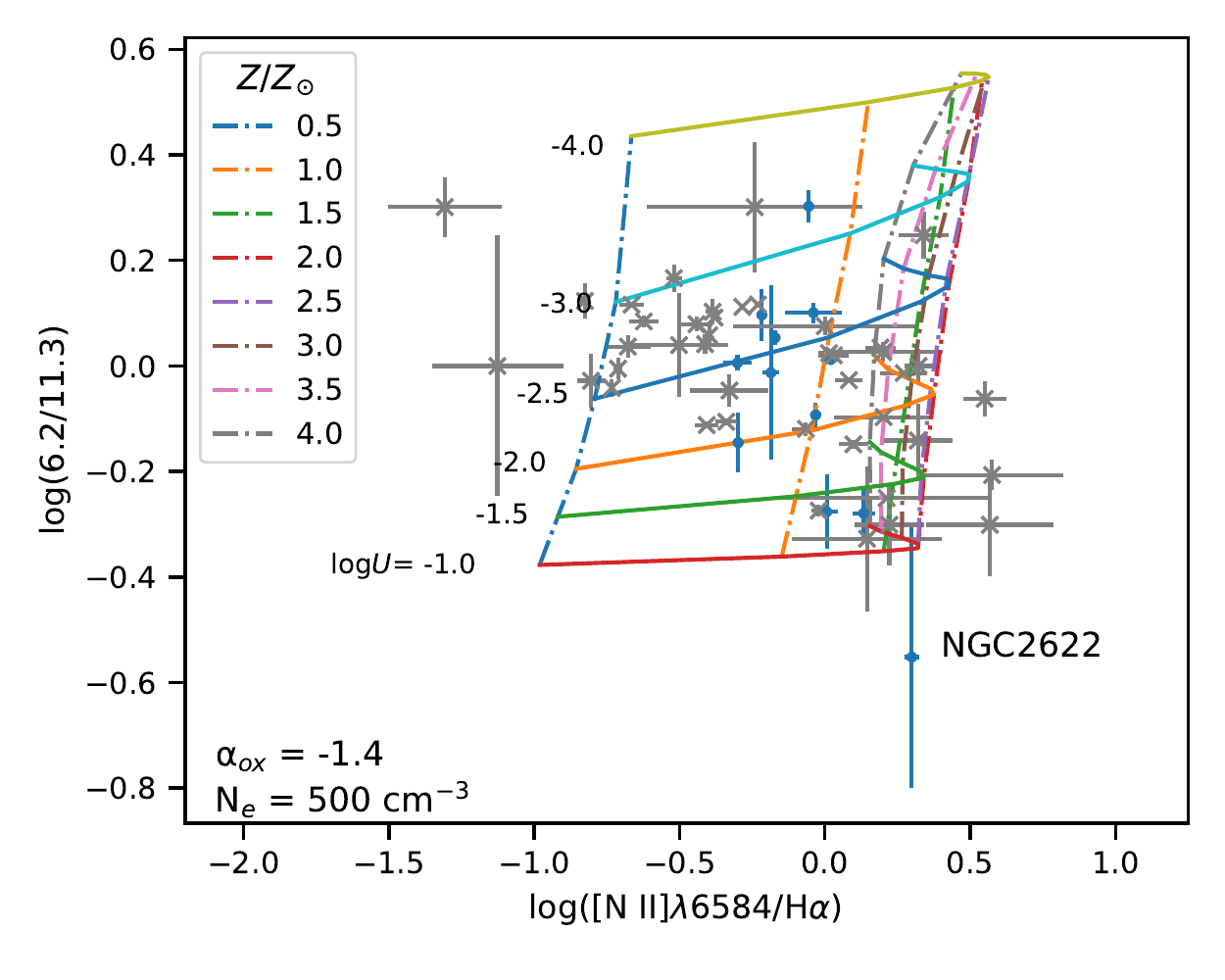}\\
\caption{log(6.2/11.3) versus log([\ion{N}{ii}]$\lambda$6584/H$\alpha$) diagram. Dashed lines connect photoionization model results  with the same metallicity, while solid lines connect models with the same logarithm of the ionization parameter $U$, as indicated. The blue points represent the  observational  line intensity ratios from the sample considered in this work, and the gray points represent data from \citet{smith2007mid}. The assumed values of the metallicity ($Z/Z_\odot$) and log\,$U$ values are indicated.}
\label{grid_photoionization_model_PAHs_NII}
\end{figure}

In Fig.~\ref{grid_photoionization_model_PAHs_NII}, a 
diagram log(6.2/11.3) versus log([\ion{N}{ii}]$\lambda$6584/H$\alpha$),  the  observational data and the photoionization model results obtained with the {\sc Cloudy} code, assuming  $\alpha_{\rm ox}=-1.4$,  are shown. It is worth to mention that [\ion{N}{ii}$\lambda$6584/H$\alpha$
line ratio   is highly sensitive to the metallicity, as measured by the oxygen abundance (O/H) and has a small (although not
negligible) ionization parameter dependence (e.g. \citealt{kewley2019, carvalho2020chemical}). An opposite behaviour is derived for the log(6.2/11.3), which shows a stronger dependence with the 
ionization degree of the gas (represented by $U$) mainly for the highest values of $U$  rather than 
with the metallicity.
We built  grid of models with  $\alpha_{ox}= -1.1$ and $-1.4$ and verify that the photoionization model results reproduce the majority of the observed line ratios assuming from the optical to X-ray spectral index $\alpha_{\rm ox}=-1.4$ (see Fig.~ \ref{grid_photoionization_model_PAHs_NII}), while for  $\alpha_{\rm ox}=-1.1$ (not shown here),  there are more points that are not reproduced by the photoionization models.  Therefore,  we adopted $\alpha_{\rm ox}=-1.4$. This value  is representative of typical AGNs, as described in the Hazy manual of the {\sc CLOUDY} code\footnote{\url{http://web.physics.ucsb.edu/~phys233/w2014/hazy1_c13.pdf}}, and  pointed out by \cite{miller2011,zhu2019} and  \cite{dors2019}. 
To calibrate the metallicity as a function of the
log(6.2/11.3), we calculated the logarithm
of the ionization parameter and the metallicity values for
each object of our sample by linear interpolations between
the models shown in Fig.~\ref{grid_photoionization_model_PAHs_NII}. 
Table \ref{tab:condições_física_estimadas} lists the oxygen abundance  and the ionizing parameter $U$ derived by  linear interpolation between the models.

\begin{table}
\caption{The oxygen abundance and the ionizing parameter $U$.}
\label{tab:condições_física_estimadas}
\centering
\begin{tabular}{lrr}
\noalign{\smallskip}
\hline
\hline
\multicolumn{1}{c}{Object} & \multicolumn{1}{c}{12 + log(O/H)$_\odot$} & \multicolumn{1}{c}{log($U$)}\\
\hline
Mrk\,273   & 8.64$\pm$0.01 & -3.42$\pm$0.11 \\ 
Mrk\,471   & 8.83$\pm$0.03 & -1.28$\pm$0.23 \\ 
Mrk\,609   & 8.69$\pm$0.02 & -2.08$\pm$0.08 \\ 
Mrk\,622   & 8.76$\pm$0.04 & -1.33$\pm$0.33 \\ 
Mrk\,883   & 8.62$\pm$0.02 & -2.00$\pm$0.33 \\ 
NGC\,660   & 8.70$\pm$0.01 & -2.37$\pm$0.05 \\ 
NGC\,2622  &   ~~~~...        &    ~~~~...      \\ 
NGC\,2623  & 8.67$\pm$0.05 & -2.62$\pm$0.10 \\ 
NGC\,4676  & 8.60$\pm$0.02 & -2.49$\pm$0.09 \\ 
NGC\,4922  & 8.63$\pm$0.00 & -2.55$\pm$0.05 \\
NGC\,5256  & 8.61$\pm$0.01 & -2.69$\pm$0.15 \\
NGC\,5347  & 8.64$\pm$0.03 & -2.35$\pm$0.61\\
\noalign{\smallskip}
\hline 
\noalign{\smallskip}
\end{tabular}
\end{table}

\section{Infrared Data Analysis}
\label{sec:data_analisis}

\subsection{{\sc PAHFIT}}
\label{subsection: analysis_ir}

The mid-IR spectrum of any object is composed of a diversity of components, such as dust and stellar continuum, emission features from PAHs, forbidden spectral lines from Ne, Ar, S, and Fe,  and rotational lines from molecular hydrogen. In low-resolution spectra, these components are blended, and it is necessary to separate them.  We employed the spectrum fitting tool {\sc PAHFIT}   \citep{smith2007mid} to decompose the  5 to 15~$\mu$m low-resolution Spitzer's IRS spectra and obtain the fluxes for different PAH features from our sample.

The default {\sc PAHFIT} parameters were adopted in the fitting. The dust continuum was represented by blackbodies at fixed temperatures of T = 35, 40, 50, 65, 90, 135, 200, and 300~K. Drude profiles were applied to recover the full strength of dust emission features, with PAH emissions at central wavelengths of 5.3, 5.7, 6.2, 6.7, 7.4, 7.6, 7.7, 7.8, 8.3, 8.6, 10.7, 11.2, 11.3, 12.0, 12.6, 12.7, 13.5, 14.0 and 14.2 $\mu$m. From these measurements, we analyzed the emissions of 6.2~$\mu$m PAH,  7.7~$\mu$m PAH complex, and 11.3 $\mu$m PAH complex. The  7.7$\mu$m PAH complex is the sum of 7.6, 7.7 and 7.8$\mu$m,  while 11.3$\mu$m PAH complex is the sum of 11.2 and 11.3 $\mu$m. 

\begin{figure}
\includegraphics[width=\linewidth]{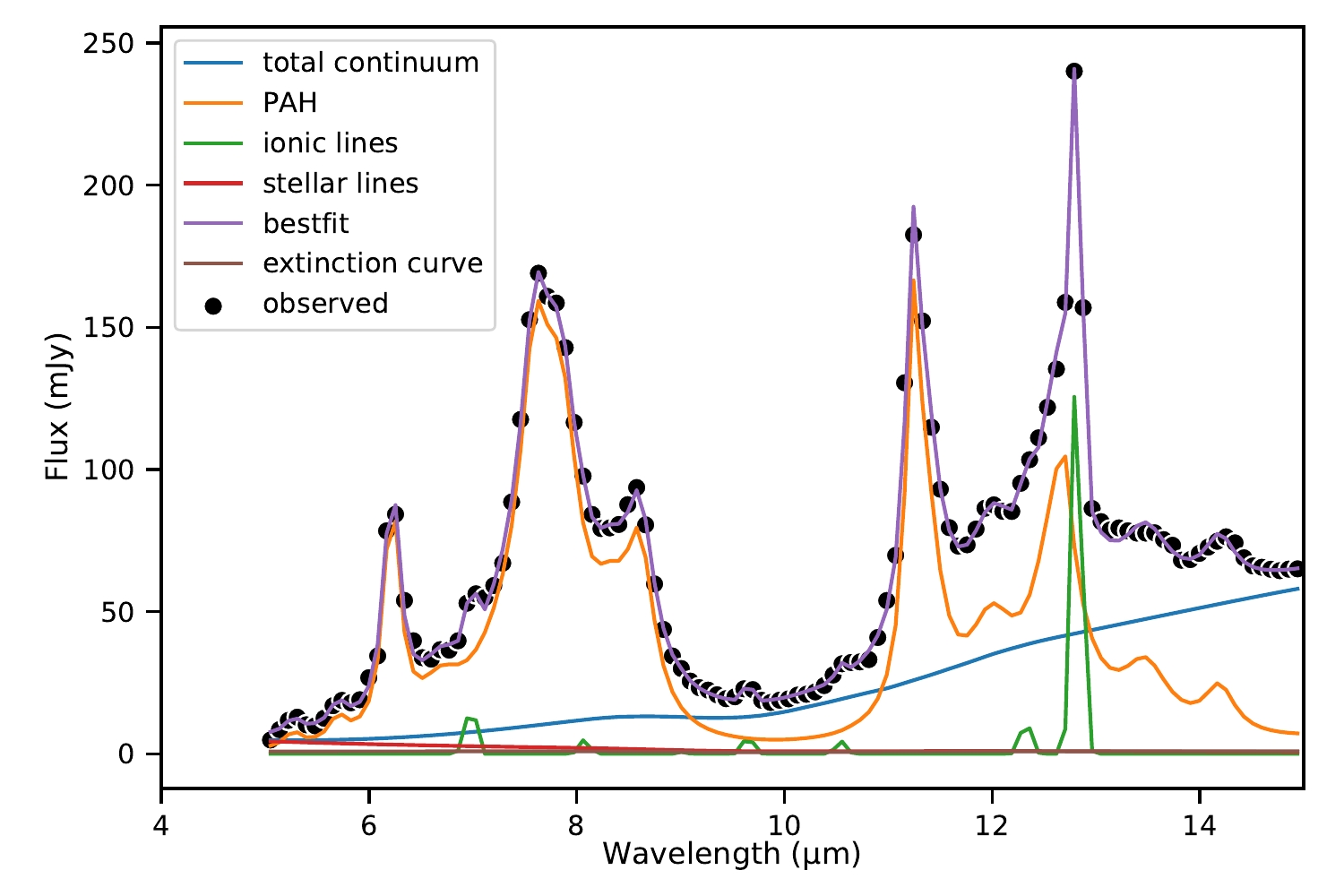}
\caption{Detailed decomposition of NGC\,4676 from 5 to 15 $\mu$m obtained with  the {\sc PAHFIT} code in components of the total continuum, ionic lines, stellar stars, best-fit model, and PAH features.}
\label{fig:spectra_decomposition_PAHFIT_ngc4676}
\end{figure}

The integrated PAH fluxes analyzed in this work are given in Table~\ref{tab:fluxes_pahs_saida_pahfit}.  Fig.~\ref{fig:spectra_decomposition_PAHFIT_ngc4676} shows the detailed decomposition of the starburst galaxy NGC\,4676 from 5 to 15~$\mu$m obtained with the {\sc PAHFIT} code in the components: total continuum, ionic lines, stellar lines, best-fit model, and PAH features. This spectral decomposition was similarly applied to the other eleven objects in our sample.

\begin{figure} 
\includegraphics[width=\columnwidth]{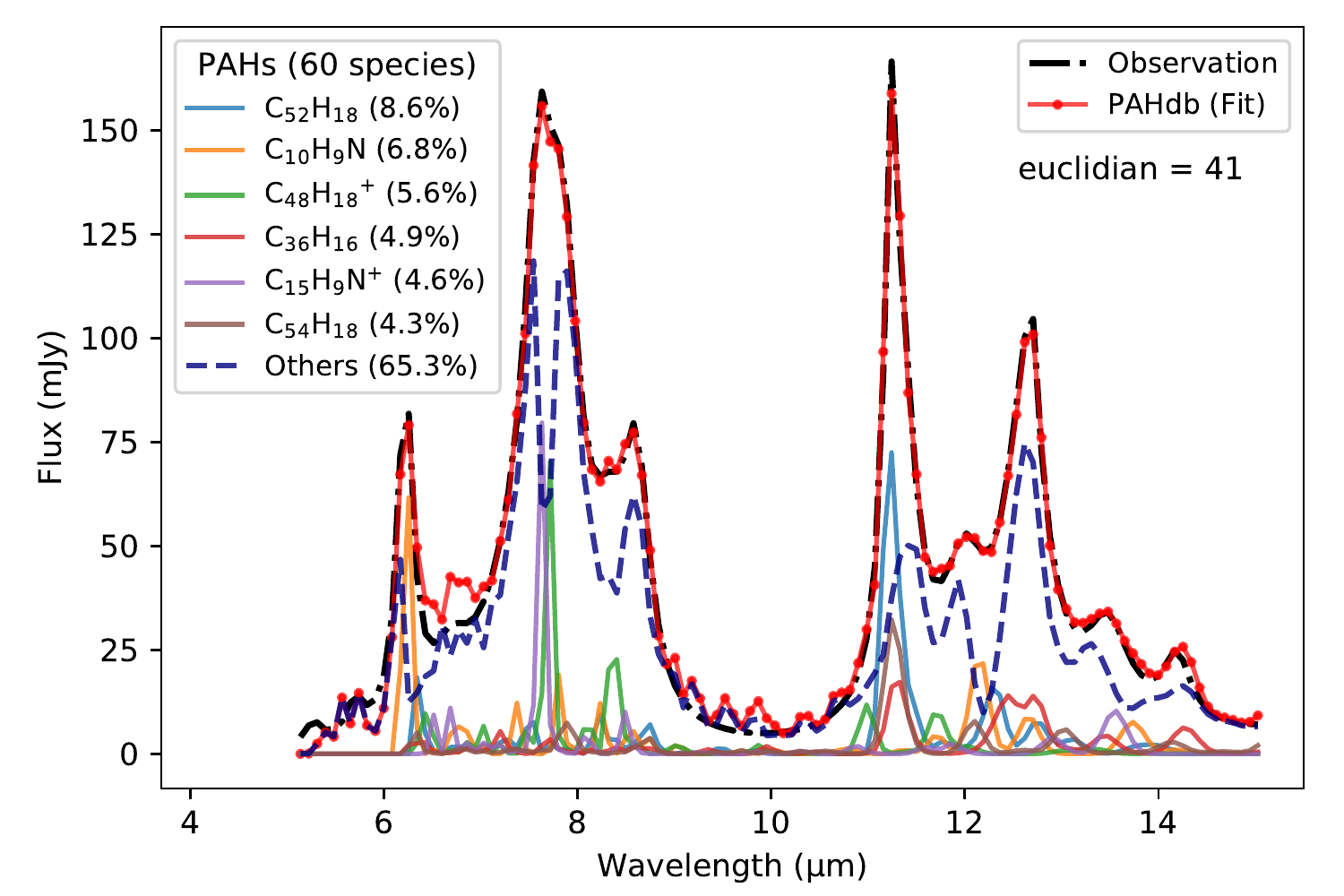} 
\caption{The PAHdb model fitting for the  observed spectrum of NGC\,4676. The quality of the fitting is quantified by the parameter euclidean. The fractional contribution to the flux from the species is shown.}
\label{fig:species_ngc4676}
\end{figure}

\subsection{PAH spectra}
The PAH spectra of our sample obtained using the {\sc PAHFIT} were analyzed using version 3.2 of the NASA Ames PAH IR Spectroscopic Database \citep[PAHdb,][]{boersma2014nasa,bauschlicher2018nasa}, which contains 4\,233 theoretically calculated PAH spectra.  In order to perform the fit, we used the version online of the PAHdb, which is based on spectra of individual aromatic molecules with specific charge states, structures, compositions, and sizes.  The  Gaussian line profile with a full width at half maximum (FWHM) of 15~cm$^{-1}$ and a uniform shift of 15~cm$^{-1}$ for all bands was assumed to mimic some effects of anharmonicity.  Fig.~\ref{fig:species_ngc4676} illustrates the fitting obtained for the spectrum of NGC\,4676, and it shows the main species that are contributing to the IR emission in NGC\,4676.  This fitting resulted in sixty species contributing to the galaxy’s flux. The contributions of the six species that contribute  most to the galaxy's flux are in parenthesis and the their adjusts are showed in different colors. In blue, the contribution of all fifty-four other species. In Fig.~\ref{fig:species_ngc4676}, it is possible to see the quality of the fit measured by Euclidean norm\footnote{Euclidean norm is one mathematical method that measures the distance between two points. In this case, the observed and the theoretical ones. Therefore, the lower the Euclidean norm, the better the quality of the fit.}, whose results are displayed at the top right of the figure. The fitting obtained for the remaining objects is shown in Figs.~ \ref{Fig:species_all_pahdb1} and \ref{Fig:species_all_pahdb2} (see Appendix B).

\section{RESULTS AND DISCUSSION }
\label{sec:results_discussion}

\subsection{Oxygen abundance and the ionization parameter}
\label{subsubsec:phisical_properties_estimated}

The effects of the radiation field and metallicity on the intensity ratios and PAH equivalent widths have
been reported by many authors in the literature. Due to the significant radiation field of AGNs  as revelead by the presence of a significant hot dust continuum, the small PAHs could be destroyed, which tend to decrease the 6.2 and 11.3 $\mu$m intensities  (e.g., \citealt{sturm2000iso,desai2007pah,diamond2010infrared}). \citet{smith2007mid} performed a study for a sample of Seyfert, LINER and star forming galaxies, and found a 
trend of the increase of the integrated luminosity of the PAH bands, relative to the total infrared, with the increasing of  the oxygen abundance. Besides, these authors verified that the 7.7/11.3  PAH intensity ratio  does not vary with the radiation hardness, measured as [\ion{Ne}{iii}]/[\ion{Ne}{ii}] for star forming galaxies, but shows a dependence for AGNs, specially for [\ion{Ne}{iii}]/[\ion{Ne}{ii}]> 0.1. However, \citet{maragkoudakis2018pahs} verified a very weak dependence of the 7.7/6.2 on the hardness of the radiation field and the ionization index for the \ion{H}{I} regions along the M\,83 and M\,33 galaxies (for a review, see for instance, \citealt{li2020spitzer}).

\begin{table*}
\centering
\caption{Integrated PAH fluxes (in units of $10^{-16}$ W/m$^{2}$) obtained by the {\sc PAHFIT} code.}
\label{tab:fluxes_pahs_saida_pahfit}
\begin{tabular}{lrrrrrrrrrrr}
\noalign{\smallskip}
\hline
\hline
\noalign{\smallskip}
Object  & 6.2 $\mu$m & 6.7 $\mu$m & 7.4 $\mu$m & 7.6. $\mu$m & 7.8 $\mu$m     & 8.6 $\mu$m   & 11.2 $\mu$m & 11.3 $\mu$m & & 12.6 $\mu$m & 12.7 $\mu$m  \\
\hline
& & & \multicolumn{3}{c}{7.7 $\mu$m Complex} & & \multicolumn{2}{c}{11.3 $\mu$m Complex} & & \multicolumn{2}{c}{12.6 $\mu$m Complex} \\ 
\noalign{\smallskip}
\cline{4-6}
\cline{8-9}
\cline{11-12}
\\

 MrK\,273 &     39.40 &     36.70 &     96.60 &     51.90 &     82.80 &     31.40 &      12.10 &       7.53 &        &      23.10 &       2.34 \\
  Mrk\,471 &      3.79 &      0.49 &      0.06 &      5.60 &      5.80 &      3.28 &       1.42 &       5.79 &        &       3.54 &       0.26 \\
  Mrk\,609 &     15.20 &      4.76 &      9.52 &     24.40 &     25.40 &     14.40 &       3.52 &      15.30 &        &       8.50 &       1.33 \\
  Mrk\,622 &      3.74 &      1.45 &      3.59 &      5.19 &      5.52 &      2.95 &       2.31 &       4.76 &        &       3.81 &       0.00 \\
  Mrk\,883 &      2.97 &      0.74 &      0.36 &      4.22 &      3.67 &      1.77 &       0.96 &       3.19 &        &       1.57 &       0.36 \\
  NGC\,660 &    263.00 &    112.00 &    434.00 &    322.00 &    496.00 &    235.00 &      39.60 &     216.00 &        &     156.00 &      23.90 \\
 NGC\,2622 &      0.59 &      0.43 &      0.00 &      1.57 &      1.53 &      0.56 &       0.58 &       1.53 &        &       0.23 &       0.22 \\
 NGC\,2623 &     46.80 &     19.20 &     93.80 &     51.10 &     86.80 &     50.80 &      12.70 &      24.40 &        &      30.50 &       4.03 \\
 NGC\,4676 &     21.60 &      7.49 &     28.30 &     27.50 &     31.50 &     15.40 &       6.51 &      14.80 &        &      12.30 &       1.59 \\
 NGC\,4922 &     16.00 &      8.02 &     21.60 &     18.40 &     21.90 &     10.10 &       4.04 &      10.10 &        &       7.13 &       1.23 \\
 NGC\,5256 &     32.10 &     11.10 &     54.30 &     41.30 &     49.10 &     23.00 &       7.18 &      18.50 &        &      19.50 &       0.54 \\
 NGC\,5347 &      7.20 &     10.00 &     28.80 &      2.63 &      8.32 &      4.79 &       0.92 &       6.49 &        &       5.99 &       0.00 \\

\noalign{\smallskip}
\hline
\noalign{\smallskip}
\end{tabular}
\end{table*}

\begin{figure*}
\includegraphics[width=0.9\linewidth]{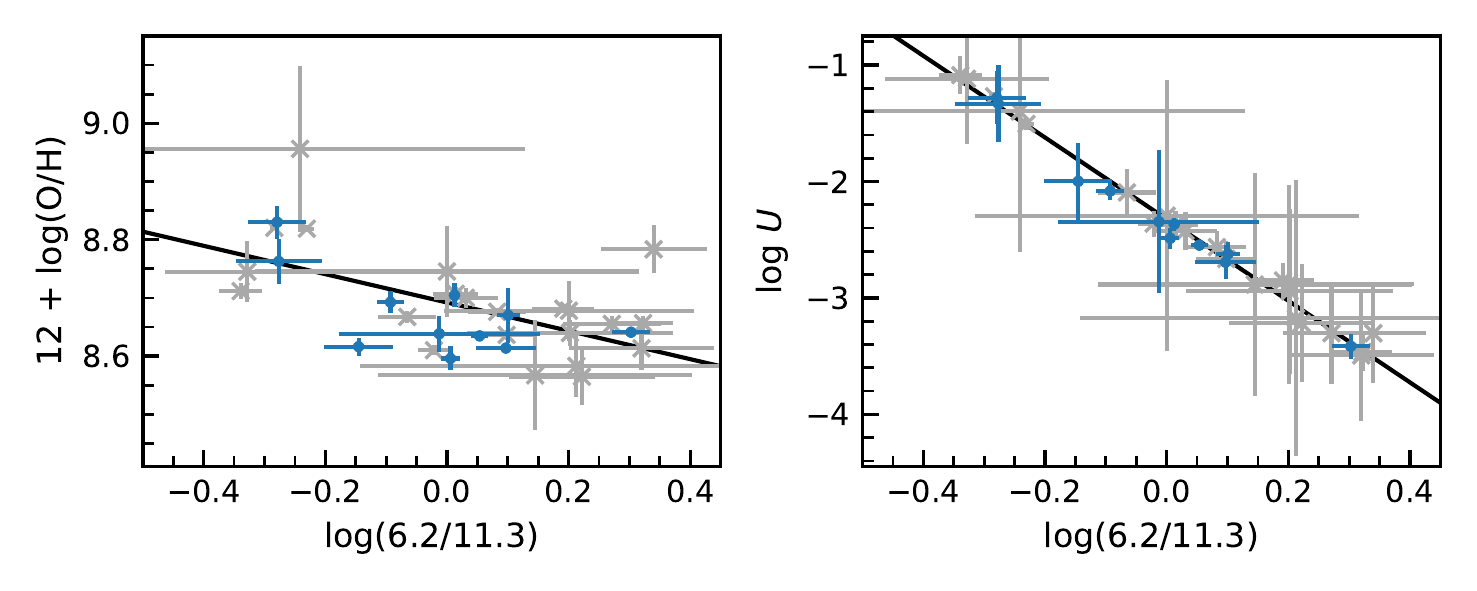}
\caption{6.2/11.3 emission ratio as a function of the  oxygen abundance (12 + log(O/H)) and the logarithm of the ionization parameter of hydrogen (log $U$). Blue points represent the galaxies of our sample and gray points the SINGs sample of galaxies from \citet{smith2007mid}. The black solid line is the linear fit to the data.}
\label{PAHs_abundance_logU_from_Cloudy}
\end{figure*}

Fig.~\ref{PAHs_abundance_logU_from_Cloudy} presents the 6.2/11.3 $\mu$m emission ratio as a function of the oxygen abundance (12 + log(O/H)) and the logarithm of the ionization parameter of hydrogen (log $U$) for our sample of galaxies and the SINGs sample by \citet{smith2007mid}. We notice that the 6.2/11.3 $\mu$m PAH intensity presents a linear correlation with the  oxygen abundance, and the ionization parameter $U$. We find the best fit line of  12+ log(O/H) = -0.24$\times$ log(6.2/11.3) + 8.69, with  $R=-0.58$, and log($U$) = -3.51$\times$ log(6.2/11.3) - 2.32, with a Pearson correlation coefficient of $R=-0.99$. It is important to emphasize, that for the first time we found that the photoionization models can reproduce most of the observational data in the log (6.2/11.3) versus log([\ion{N}{ii}]$\lambda$6584/H$\alpha$) diagram.

\subsection{Species Identification and Degeneracy}

The knowledge on the physico-chemical conditions of the environments where PAHs and PANHs exist is of great importance to understanding the molecular formation and destruction, with the possibility of new compounds being formed. Many issues have been raised about which aromatic species can be found in space environments, while theoretical studies have been conducted out to determine vibration frequencies of potential candidates \citep{ricca2019polycyclic,mattioda2020nasa}. Also, understanding the link between the evolutionary stage of environments and the aromatic species in them is an enigma that must be studied.  In this work, we will call PANH any aromatic molecule composed of carbon, hydrogen and nitrogen, whether nitrogen is inside or outside the ring.

The PAH abundances (PAH \%) for each species, i.e., the percentage  of the contribution on the flux of each species to the total flux in the spectral range between 5$-$15~$\mu$m, were obtained  by dividing the flux contribution  of each species by the total flux, which includes the contributions of all species. The total number of species that contribute to IR emission to each galaxy varies from 49 to 77. These species are presented in Appendix~\ref{appendix:tab_species_emission_each_galaxy} for the twelve studied objects. The most relevant identified species in the sample contain 10 $-$ 82 carbon atoms, while the overall species can be composed of 9 $-$ 170 carbon atoms.

Among the pure hydrocarbons, we highlight the  C$_{52}$H$_{18}$ (UID\,3173), which is present in 75 per cent of the sample and contributes from about 4 per cent to 13 per cent in the galaxy's fluxes. Among nitrogenous polycyclic aromatic hydrocarbons, C$_{10}$H$_{9}$N (UID\,473) is present in 92 per cent of the sample, with fluxes contributions ranging from about 3 to 8 per cent. 
The molecular ion C$_{10}$H$_{9}$N$^{+}$ is also present in several galaxies analyzed in this work. 
Similar species such as benzonitrile, also called cyanobenzene (C$_{6}$H$_{5}$CN), and cyanonaphthalenes (C$_{10}$H$_{7}$N) have already been detected in the dark molecular cloud TMC-1 \citep[]{McGuire18,mcguire2021detection, McCarthy21}. Confirmation of hydrocarbons with CN, such as cyanobenzene in TMC-1 and other sources, improves our understanding of chemical processes. Reactions between unsaturated hydrocarbons with CN are usually exothermic and without barrier activation. Furthermore, CN is an abundant radical in most molecular clouds. Since cyan derivatives have significant dipole moments, they are easy targets to detect in radio and, therefore, will have bright rotation spectra \citep[]{McCarthy21}.

 It is known that there is degeneracy in the fitting of the spectra using the  PAHdb database, and it is not possible to guarantee that these three species are always present in the spectra of the galaxies studied here. If C$_{52}$H$_{18}$, C$_{10}$H$_{9}$N, and C$_{14}$H$_{11}$N are excluded from the database, the PAH spectra can still be reproduced, but the quality of the fitting decreases and the Euclidean have a slightly worse result. 
In this context, it is worthwhile to known that the species C$_{14}$H$_{11}$N,  is simply C$_{10}$H$_{9}$N with the addition of one more aromatic ring. This molecule is present in 60 per cent of the sample, and contributes of  2 to 6 per cent of the total flux.  For example, the fitting of the spectrum  of the galaxy Mrk\,273  performed using all database  resulted in the contribution of 56 species (see the fitting in  Fig.~\ref{Fig:species_all_pahdb1}), with an Euclidean close to 95. When the species C$_{52}$H$_{18}$, C$_{10}$H$_{9}$N, C$_{14}$H$_{11}$N and their isomers (15 species) were removed, the PAHdb resulted in 53 species contributing to the galaxy's flux and the Euclidean value was 108.  In this  case, some new species were introduced, as well as the relative quantity of the species was slightly modified, when compared with the first fitting (considering all database). 
However,  these new species  have the same characteristics of the  excluded species, being, majority, small pure PAHs or small PAHs with heteroatom (N or O), with two or three fused aromatic rings.  Therefore, although excluded molecules are more appropriate, molecules of the same family also reproduce the spectra of the galaxies, and  the presence and (relative) quantity of the PAH  species is not uniquely determined.

Then, a coarse, but robust description of the contribution of the types of PAHs of a galaxy may be obtained by grouping their different types. We grouped
these types adapted according to \citet{andrews2015pah}, as dehydrogenated, pure, PANHs, and heteroatom. The table ~\ref{tab:out_pahdb_agrupado} lists the results of flux contribution according to these components.

\begin{table}
\centering
\caption{Flux contribution  grouped by type of PAHs.}
\label{tab:out_pahdb_agrupado}
\begin{tabular}{lp{0.18\linewidth}ccc}
\noalign{\smallskip}
\hline
\multicolumn{1}{c}{Object} & Dehydrogenated  & Pure PAHs & PANHs & Heteroatom \\

& (\%) & (\%)& (\%) &(\%) \\
\hline
\hline
Mrk 273                    & 42.1                & 25.1      & 32.5  & 0.3        \\
Mrk 471                    & 31.5                & 53.3      & 11.9  & 3.4        \\
Mrk 609                    & 32.2                & 55.3      & 10.4  & 2.0        \\
Mrk 622                    & 36.2                & 48.9      & 10.8  & 4.2        \\
Mrk 883                    & 28.6                & 60.0      & 8.3   & 3.1        \\
NGC 660                    & 33.8                & 44.3      & 19.0  & 3.0        \\
NGC 2622                   & 37.3                & 51.1      & 10.8  & 0.8        \\
NGC 2623                   & 35.2                & 41.4      & 23.4  & 0.0        \\
NGC 4676                   & 30.2                & 48.9      & 19.9  & 1.1        \\
NGC 4992                   & 32.0                & 49.9      & 15.9  & 2.2        \\
NGC 5256                   & 31.6                & 48.6      & 19.4  & 0.4       \\
NGC 5347                   & 47.8                & 36.2      & 15.1  & 1.3        \\
\hline
\noalign{\smallskip}
\end{tabular}
\end{table}

\subsection{PANHs and the Peeter Classification}


\citet{hudgins2005variations} demonstrated that only the nitrogen incorporated into the aromatic rings is capable of reproducing the interstellar observations of the 6.2~$\mu$m PAH band at shorter wavelengths through experimental process.
The blueshift of the 6.2~$\mu$m band was also observed in 67 per cent of 155 starburst-dominated galaxies \citep{canelo2018variations}.
Moreover, \citet{boersma2013properties} proposed that the PANH cations dominate both the 6.2 and 11.0~$\mu$m emissions in slightly dense regions of the Photodissociation Region (PDR) known as NGC\,7023, and they are also significantly responsible for the emissions of the 7.7, 8.6, 12.0, and 12.7~$\mu$m features.

 PANHs, both neutral and cation, generate up to 55 per cent  of emission at 6.2 $\mu$m in our sample of galaxies. PANHs also contribute, on average,  about 5.7 per cent of the band flux in 11.3 $\mu$m, reaching the value of 37 per cent in the MRK\,883. 

The importance of N-containing PAHs in the total emission of our sample is also in agreement with the results obtained by  \citet{canelo2018variations} and \citet{Canelo2021}, as can be seen in Table \ref{tab:carla-classes}. This table shows the Peeters’ classification of the 6.2, 7.7, and 8.6~$\mu$m bands \citep{peeters2002rich}, previously obtained by \citet{thesis-canelo}
to most of our galaxies together with our current classification for NGC\,2622, NGC\,4922 and NGC\,5347.  The procedure for this analysis, as well as the new results from those three sources, are described in Appendix~\ref{ap:peeters}. The 6.2~$\mu$m band, in particular, was mainly distributed into the A class objects, implying a predominance of blue shifted profiles, which are typical of PANH emission and thus confirms the presence of such molecules in these sources.

Besides, we found that the majority of the PAH population is composed of up to 95 per cent of small species\footnote{In the present work, small PAHs are considered to have a number of carbon atoms lower than 50 (N$_{C}$~<~50), while large PAHs contain N$_{C}$~$\geq$~50 atoms.} (81 per cent being the average) and 79 per cent of neutral PAHs ( 68 per cent the average) in our sample of galaxies. These values are consistent with those  provided by \citet{martins2021starbursts}, who found that the predictions by the code PAHdb species of a sample based on Spitzer/IRS mid-infrared spectra of ~700 dusty galaxies from the GOALS and ATLAS surveys shown themselves to be predominantly small and neutral PAHs. 

In this sense, we can attribute a young object profile to our studied sources with ISM-type environments and a PAH population dominated by small species \citep{shannon2019examining}.  Therefore, the spectra of our sample are consistent with Peeters' A class object, which corresponds to \ion{H}{ii} regions and the general material illuminated by a star in the ISM. On the other hand, large PAHs are normally more connected with B class spectra of evolved objects, such as circumstellar material, planetary nebula, and a variety of post-AGB (asymptotic giant branch) stars  \citep[e.g][]{tielens2008interstellar, andrews2015pah, peeters2017pah}. Moreover, the B class profiles could also be a mixture of PANH, small and large PAH emissions \citep{peeters2002rich}. It is important to note that \citet{Yang2017} suggested that neither PAHs with a large side chain nor PAHs with unsaturated alkyl chains are expected to be present in the ISM in a large abundance.  From  Table~\ref{tab:carla-classes}, both A and B classes are present in our sample, which emphasizes the importance of spatially-resolved observations in the determinations of which locations in the galaxies where those different PAH populations are more relevant.


Furthermore, \citet{boersma2013properties} and \citet{andrews2015pah} showed that the dominance of the emissions of small molecules increases with the proximity of the ionizing source. \citet{draine2001infrared} also demonstrated this behaviour when they proposed that small PAHs emit mainly at 6.2 and 7.7 $\mu$m bands, while large PAHs emit at longer wavelengths \citep{allamandola1999modeling}. Concerning the neutral species, their predominance is also expected in photodissociation regions where the species can be shielded due to high density/low temperature environments. In fact,  \citet{allamandola1999modeling} demonstrated that the absorption spectrum produced by the neutral PAHs, compared to the spectrum produced by the same PAHs in the cationic form, segregate separately in different regions of the IR when these PAHs are small. According to \citet{bauschlicher2008infrared},  small PAHs emit at 5~--~9~$\mu$m region while the emissions at shorter wavelengths of the 11.3~$\mu$m profile would be produced by neutral PAH molecules.

\begin{table}
\centering
\caption{Peeters' classification of the 6.2, 7.7 and 8.6~$\mu$m PAH bands of the studied galaxies.}
\label{tab:carla-classes}
\begin{tabular}{lcccc}
\hline
Source   & 6.2 $\mu$m & 7.7 $\mu$m & 8.6 $\mu$m  & Reference \\
  & Class & Class & Class & \\
\hline
Mrk\,273  & A & B & A & [1,2] \\
MrK\,471  & B & A & B & [1,2] \\
Mrk\,609  & A & B & B & [1,2] \\
MrK\,622  & A & A & A & [1,2] \\
Mrk\,883  & B & A & A & [1,2] \\
NGC\,660  & A & A & A & [1,2] \\
NGC\,2622 & B & A & A & This work \\
NGC\,2623 & A & B & B & [1,2] \\
NGC\,4676 & A & B & A & [1,2] \\
NGC\,4922 & A & A & B & This work \\
NGC\,5256 & A & A & B & [1,2] \\
NGC\,5347 & B & C & ... & This work \\
\hline
\end{tabular} \\
References: [1] \citet{canelo2018variations}; [2] \citep{Canelo2021}. 
\end{table}

\subsection{Small and ionized PAHS, PANHS and the ionization parameter}


 It is widely  known that the size of PAH species and their ionization states can directly be related to the physical conditions of the local environment \citep{tielens2005physics,sidhu2020principal}, in specially of the ionizing field, and a factor of its characterization is the ionization parameter $\log U$. 

\begin{figure*}
\begin{minipage}{.49\textwidth}
\centering
\includegraphics[width=0.9\textwidth]{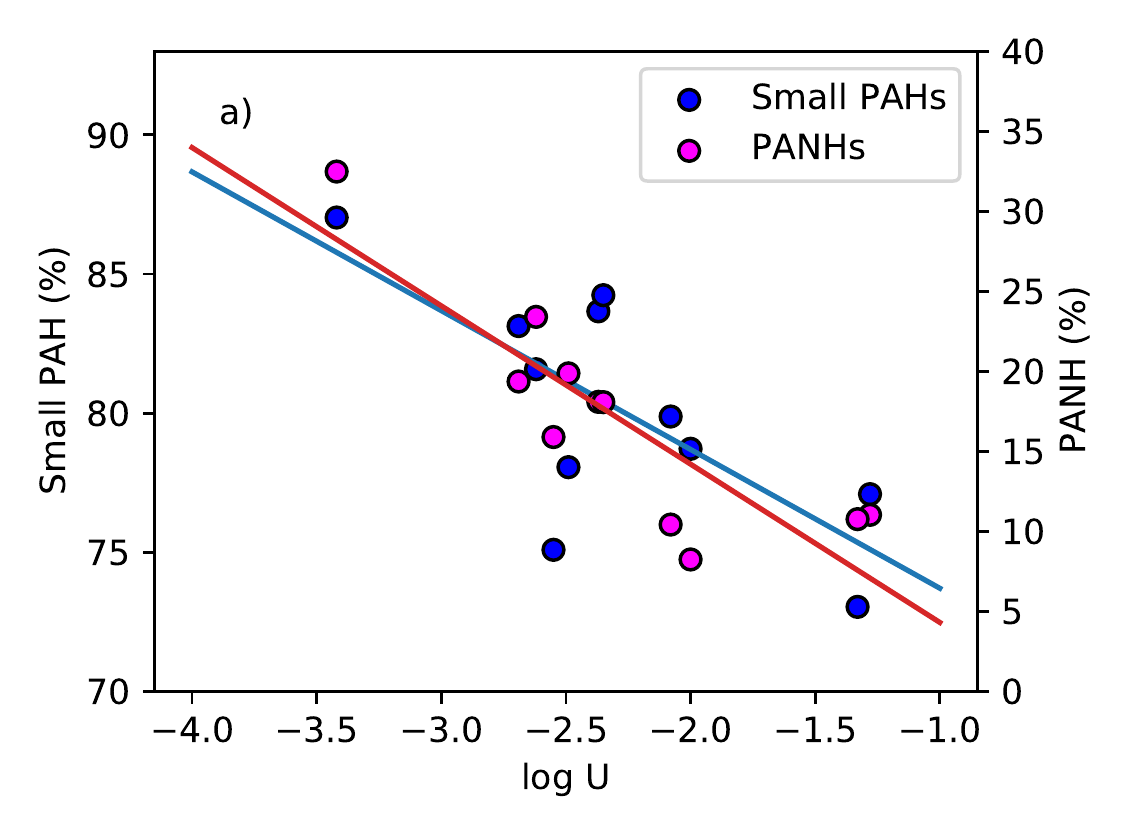}
\end{minipage}
 \begin{minipage}{.49\textwidth}
 \includegraphics[width=0.9\textwidth]{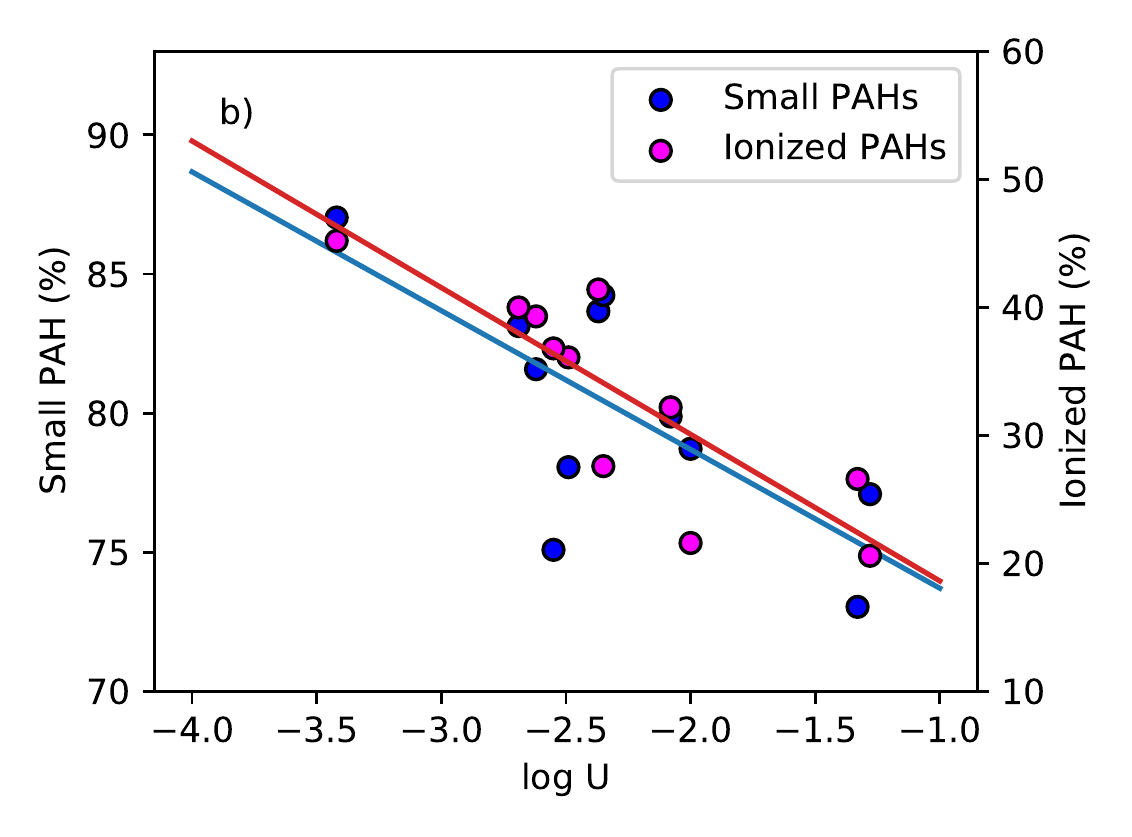}
 \end{minipage}
 \caption{Small PAHs and  PANHs  versus  log $U$ (right) and small PAHs versus log $U$  (left). The red and blue solid lines represent the best fits between the percentage total flux of the species (small, ionized PAHs and PANHs) and log$U$.}
 \label{graf:small_ionized_PANH_logU}
 \end{figure*}

In this section, we investigate the relation among the contribution of small PAHs, PANHs and ionized PAHs and  log $U$. In Fig.  ~\ref{graf:small_ionized_PANH_logU}, we plot these PAH populations as a function of ionization 
parameter.  We can see that the contribution of the small PAHs and PANHs drops  as the hydrogen ionization parameter increases (left panel). The correlation is very strong in both cases; with a Pearson coefficient of $R=-0.71$ (P-value$=0.012$) for the PAHs, while the PANHs has 
a $R=-0.86$ (P-value$=0.0006$). The best linear fit for the former molecules is Flux$_{(Small PAHs)}$ = -4.9 log$(U)$ + 68.76, whereas for the latter ones is Flux(PANHs) = -9.9 log $U$ - 5. The main reason for a decrease of the fraction of smaller PAHs and PANHs  with increasing ionization factor is due to the fact that the latter factor is directly correlated with the rising of the fragmentation rate of the smaller PAHs and PANHs\footnote{Most of PANHs found in our sample have N$_{C}$~<~50, except for NGC\,5347, MRK\,622 and NGC\,4676  have the isomers 
C$_{52}$H$_{18}$N$_{2}^+$ (ID=258),  and
C$_{52}$H$_{18}$N$_{2}^{2+}$ (ID=257), but all them  have contributions $\lesssim 1\%$.}, since they have lower energy of dissociation than larger PAHs \citep[e.g.,][]{peeters2005a}. For example the dissociation energy for the formation of  the cation C$_{3}$H$_{3}$ from the smaller PAH Naphthalene (C$_{10}$H$_{8}$) and aromatic hydrocarbon Benzene (C$_{6}$H$_{6}$)  is 19 eV and 14 eV, respectively (NIST webbook\footnote{\url{https:$//$webbook.nist.gov/chemistry/}}). In addition, the dissociation energy is lower for PANHs than for smaller PAHs (e.g., for dissociation of aromatic hydrocarbon 2-Methylpyridine, C$_{6}$H$_{7}$N, in the cation C$_{5}$H$_{6}$  the energy is 12.87 eV), and as most 
of PANHs in our sample are amides (an amide group outside the ring) the dissociation energy is even lower (e.g., for dissociation of aromatic  hydrocarbon Aniline, C$_{6}$H$_{7}$N, in the cation C$_{5}$H$_{6}$ the mean energy is 11.96 eV); this explains why the fraction of the 
smaller PANHs  fall ($\sim$35\% to $\sim$5\%) faster than the PAHs  ($\sim$90\% to $\sim$75\%). Likewise, the lower percentage of PANHs (mean value of  17\%) with respect to small PAHs (81\%) could be explained by their higher rates of dissociation reducing their  half-life time \citep{peeters2005a}. Therefore the fraction of small PAHs and PANHs fall as the U factor increases.

According to \citet{o2009polycyclic}, an AGN source preferentially destroys small PAH species with its strong radiation field, favouring the predominance of large species. As a consequence, the environment is more metallically enriched due to the high destruction of small PAHs. The studies by \citet{sales2010polycyclic} suggested that Seyferts tend to present an IR emission dominated by PAH molecules with more than 180 carbon atoms, while other objects such as Starbursts normally form molecules with less than 180 carbons.  Indeed, the  torus can also explain the dominance of the smaller PAHs in our sample, once it can protect the ISM-type environments of galaxies of our sample of the harder radiation field of the central engine. As can be noted, despite of the presence of the AGN these galaxies were classified as starburst-dominated sources, in which class A and B profiles and, consequently, smaller PAHs seem to be presented \citep[e.g.][]{Canelo2021,sales2010polycyclic,sales2013polycyclic}.

 The right panel of Fig.~\ref{graf:small_ionized_PANH_logU} also shows that the fraction of ionization of the PAHs decreases as the ionization factor increases. The correlation coefficient is $R=-0.84$ (P-value $=0.001$), and
the best liner fit is given by Flux(ionized PAHs) = -11.4 log $U$ + 7.2. This  result could be explained whether the PAH fragmentation rate is greater than the ionization rate, then, in this case, the percentage of ionized small PAHs diminishes as the U factor increases.
Moreover, \citet{jochims1994size}
showed that PAHs containing less than 30-40 carbon atoms, when photoexcited, tend to dissociate rather than relax by infrared emission. In the case of larger PAHs, the main relaxation way will occur via infrared emission \citep[see also][]{sales2010polycyclic,sales2013polycyclic}. 

\begin{figure}
\begin{minipage}{.49\textwidth}
\includegraphics[width=1\textwidth]{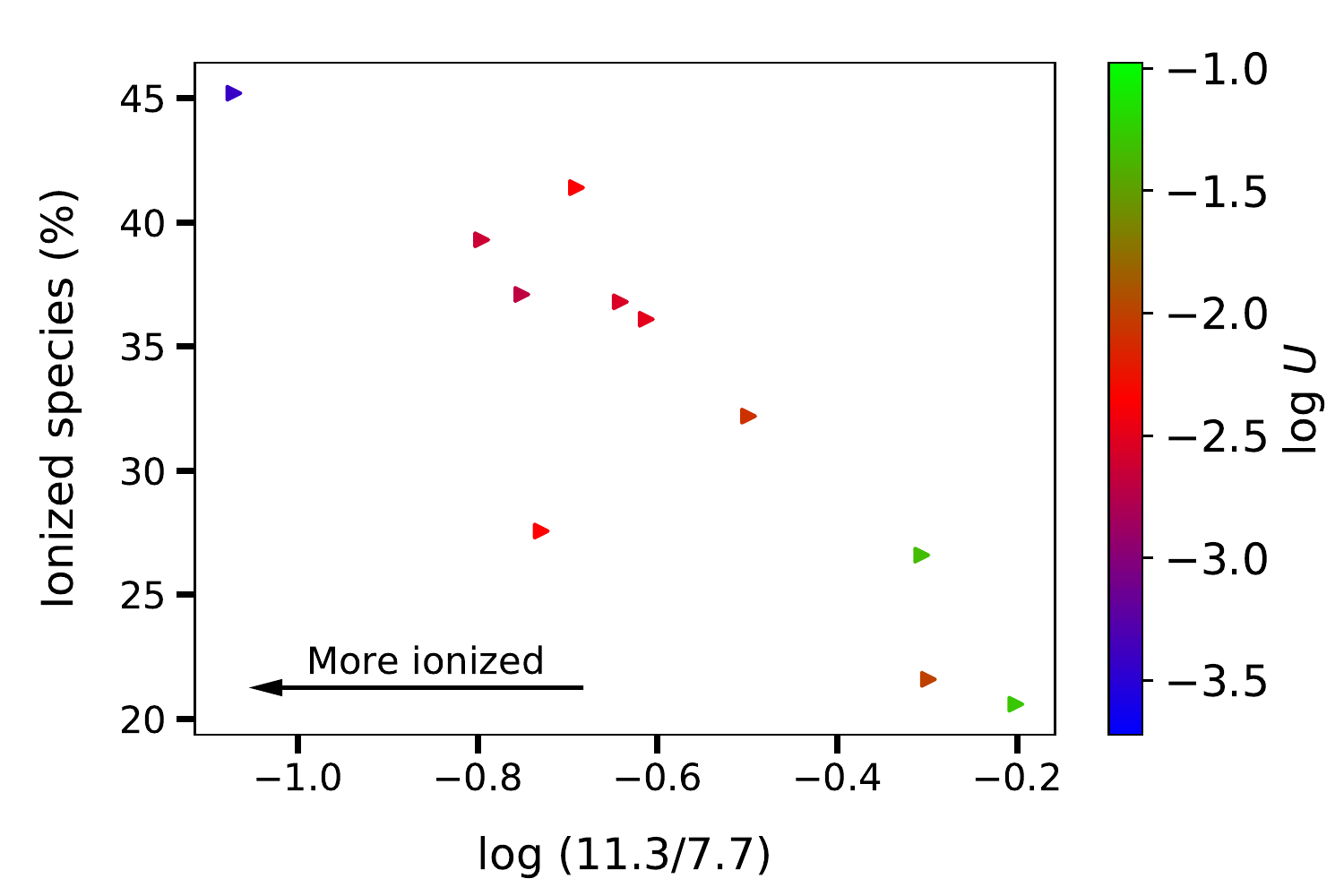}
\end{minipage}
\caption{The 11.3/7.7 emission ratios versus ionized species in percentage and log $U$.}
\label{figure1}
\end{figure}

In order to verify whether our results agrees with the classical models from {\bf \cite{draine2001infrared}}, we verified the relation between the percentage of ionization of PAHs determined from the PAHdb  with the 11.3/7.7 PAH band strength (observational) and the ionization parameter determined from the photoionization models. It is showed in  Fig.~\ref{figure1}, and as expected, we see that the ionization degree of PAH species increases with the decreasing of the 11.3/7.7 ratio and the  ionization parameter.
We found that the Ionized species (\%) $= -28.11\log(11.3/7.7) + 16.22 $, with $R = -0.90$ (P-value=0.0002).

\section{Conclusions}
\label{sec:conclusions}

In this study,  we combined optical and infrared data of a sample of Seyfert (Sy) and Starburst (SB) galaxies to characterize the main type of PAH molecules present in these objects and the local physical conditions of their irradiating sources, as well as the characteristics of the residing ionised gas. The infrared data were selected from ATLAS Spitzer/IRS with low-resolution spectra in the 5-15 $\mu$m wavelength range and local redshift, and the optical spectra from  the Sloan Digital Sky Survey, DR16, in the wavelength range of 3\,600-10\,400 \AA.
Photoionization model grids were built using the  {\sc CLOUDY} code  version 17.02 in order to compare the observational emission-line intensity ratios for the galaxies in our sample with the photoionization model predictions for similar emission-line intensity ratios. The main findings are summarised in the following bulletin:

\begin{enumerate}
    
  \item Species containing 10 $-$ 82 carbon atoms are the most abundant in the sample, while the overall species can be composed of 9 $-$ 170 carbons; 
    \\
    \item     Depending on the galaxy in our sample, the PAH population can contain up to 95 per cent of small species (the average being 81 per cent) and 79 per cent of neutral PAHs (the average  being 68 per cent);
    \\
    \item Among the pure hydrocarbons, C$_{52}$H$_{18}$ deserves the most attention, since it is found in 75 per cent of the sample, contributing from about 5 per cent to 13 per cent in the galaxy's fluxes; 
\\
    \item Concerning  PANHs, these species contribute  about 8 $-$ 32 per cent of the total emission in all the studied objects (17 per cent on average).  Aromatic amides, such as C$_{10}$H$_{9}$N, C$_{10}$H$_{9}$N$^{+}$, C$_{14}$H$_{11}$N and, C$_{15}$H$_{9}$N$^{+}$ are among the most relevant species contributing to the IR emitting flux. C$_{10}$H$_{9}$N is an important species, which is present in 92 per cent of our sample. 
    \\
    \item  Although it is not possible to guarantee that the aromatic amides mentioned above and C$_{52}$H$_{18}$ are always present in the spectra of the galaxies studied here, due to degeneracy, the removal of these molecules showed a worsening in the fit and the addition of molecules from the same family (small PAHs and aromatic amides, plus a small addition of Oxy-PAHs), suggesting that small PANHs are responsible for a large part of the emission of the galaxies in the sample. Thus, we suggest that aromatic amides, as well as  theirs cations should be considered in future experimental/theoretical studies and observations, intending to search for their interstellar detection at radio wavelengths;    
   \\
    \item For the first time, we show the AGN photoionization models  with $\alpha_{\rm ox}=-1.4$, reproducing very well the observed data points in the diagram  log(6.2/11.3) versus log([\ion{N}{ii}]$\lambda$6584/H$\alpha$);
    \\
    \item The 6.2/11.3 PAH intensity ratio presents 
a linear anti correlation between the oxygen abundance and log $U$, in the sense that the 
 6.2/11.3 ratio decreases as the oxygen abundance and log $U$ increases;
  \\
\item In our sample, the small PAH flux, as well as the ionized PAH and PANH fluxes, exhibit a decrease trend as hydrogen ionization parameter increases. Notice a linear anti correlation between these parameter. We suggest that this bevahior is directly correlated with the rising of the fragmentation rate of the smaller PAHs and PANHs at high U, since they have lower energy of dissociation than larger PAHs. In addition, the dissociation energy is lower for PANHs than for smaller PAHs, according to the observed behaviour, where the fraction of the smaller PANHs  fall (~35\% to ~5\%) faster than the PAHs  (~90\% to ~75\%) at the considered log(U) range. On the other hand, the 
decreasing of ionized PAHs with the increasing of U factor could be explained whether the PAH fragmentation rate is greater than the ionization rate.

\item {We found  that the ionization degree of PAH species increases with the decreasing of the 11.3/7.7
ratio and the  ionization parameter (log $U$), in agreement with the models proposed by {\bf \cite{draine2001infrared}}.

}
\end{enumerate}  


\section*{Acknowledgements}
A.Silva-Ribeiro acknowledges support from Coordenação de Aperfeiçoamento de Pessoal de Nível Superior (CAPES). 
A.C.K acknowledges support from Conselho Nacional de Desenvolvimento Cient\'ifico e Tecnol\'ogico (CNPq) and   Funda\c c\~ao de Amparo \`a Pesquisa do Estado de S\~ao Paulo (FAPESP), 
process number 2020/16416-5. CMC acknowledges the support of CNPq, process number 141714/2016-6, and CAPES - Finance Code 001. D. P. P. Andrade acknowledges support from CNPq and Funda\c c\~ao de Amparo \`a Pesquisa do Estado do Rio de Janeiro (FAPERJ).  D. A. Sales acknowledges support from CNPq and Funda\c c\~ao de Amparo \`a Pesquisa do Estado do Rio Grande do Sul (FAPERGS).  J.A.H.J. acknowledges support from FAPESP, process number 2021/08920-8. 


\section{Data Availability}
The data underlying this article will be shared on reasonable request to the corresponding author.

\bibliographystyle{mnras}
\bibliography{bibliografia} 


\appendix



\section{{\sc PAHFIT} spectra decomposition}
\label{appendix:decomposition_PAHFIT}

\begin{figure*}
\includegraphics[width=0.9\linewidth]{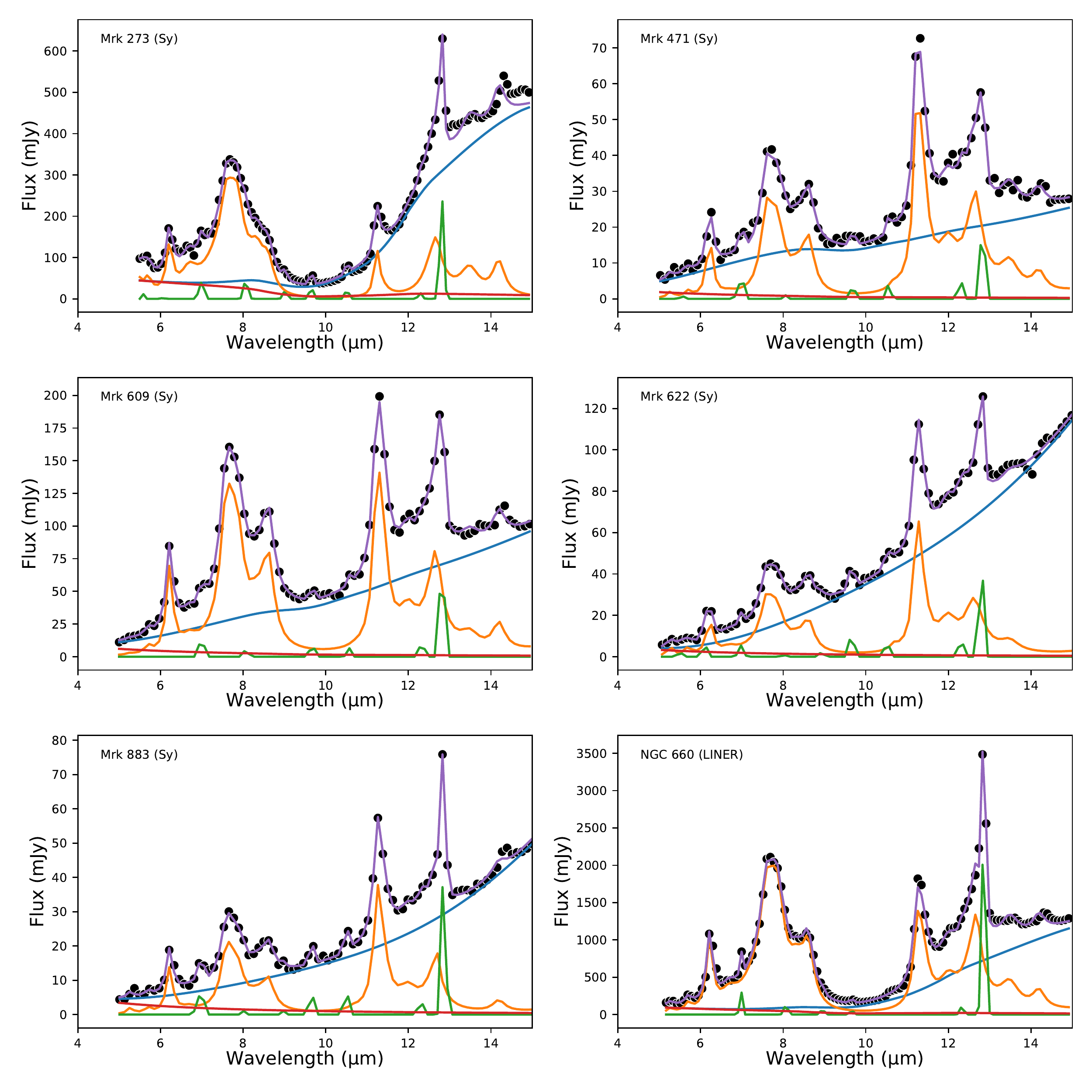}
\caption{{\sc PAHFIT} decomposition of our galaxies from 5 to 15~$\mu$m in components of the total continuum (blue line), stellar continuum (red line), ionic lines (green line), PAH and dust features (orange line) and bestfit model (purple line). The data points are shown in black.}
\label{fig:pahfit_part1}
\end{figure*}

\begin{figure*}
\includegraphics[width=0.9\textwidth]{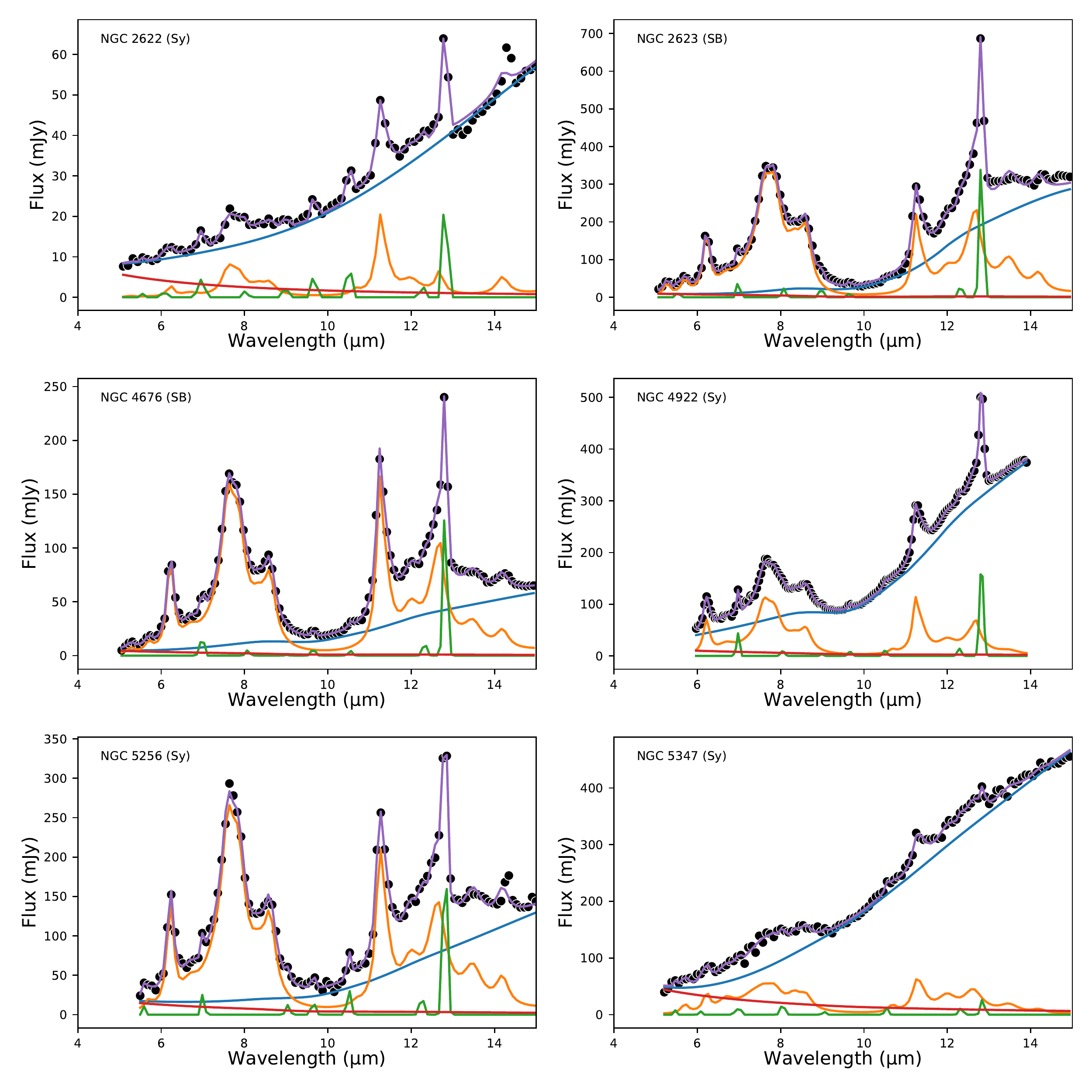}
\caption{{\sc PAHFIT} decomposition of our galaxies from 5 to 15~$\mu$m in components of the total continuum (blue line), stellar continuum (red line), ionic lines (green line), PAH and dust features (orange line) and bestfit model (purple line). The data points are shown in black.}
\label{fig:pahfit_part2}
\end{figure*}

\section{PAHdb fitting}
\label{anexo_padhdb}

\begin{figure*}
\includegraphics[width=0.88\textwidth]{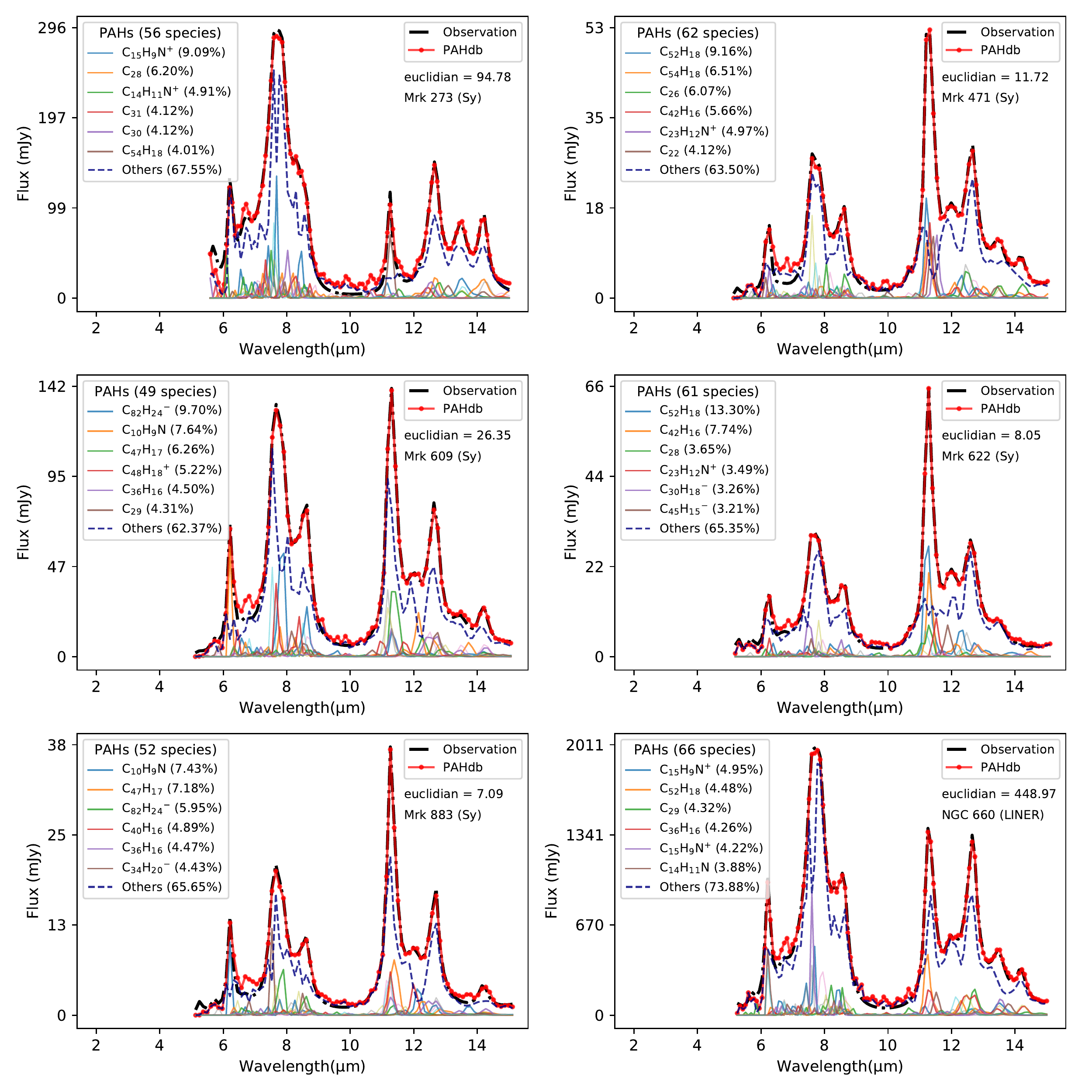}
\caption{The PAHdb fit to the  observed spectrum for the remaining of the objetcs of our sample. The quality of the fitting is quantified by the parameter euclidean. The fractional contribution to the flux from the species is shown.}
\label{Fig:species_all_pahdb1}
\end{figure*}

\begin{figure*}
\includegraphics[width=0.9\linewidth]{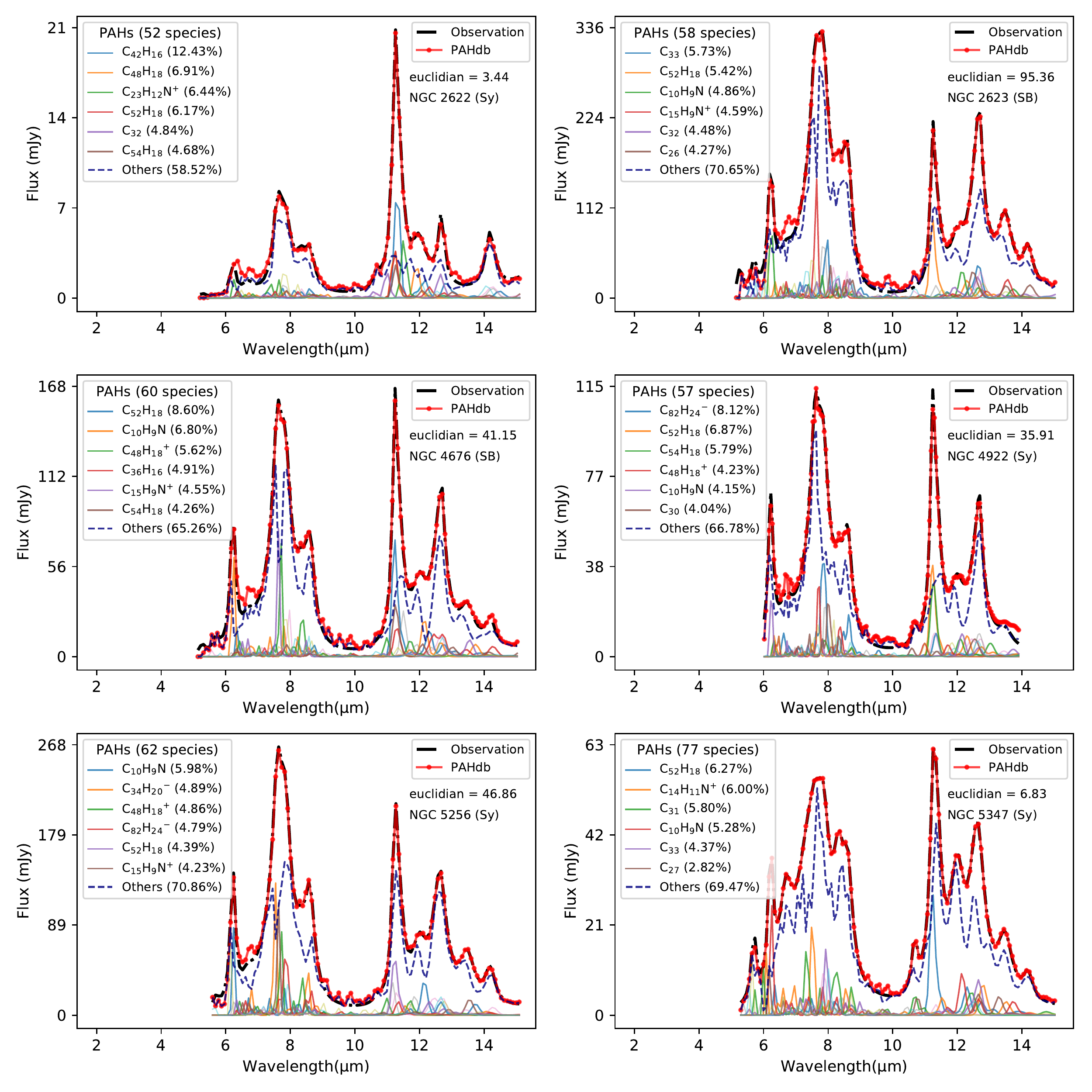}
\caption{The PAHdb fit to the  observed spectrum for the remaining of the objetcs of our sample. The quality of the fitting is quantified by the parameter euclidean. The fractional contribution to the flux from the species is shown.}
\label{Fig:species_all_pahdb2}
\end{figure*}

\section{Species emissions each galaxy}
\label{appendix:tab_species_emission_each_galaxy}

\vspace{-10cm}

\begin{table}
\centering
\caption{Species of PAHs that are contributing to the total emission in  IR spectrum of Mrk\,273. The full table is available online.}
\label{tab:emission_species_mrk273_Sy}
\begin{tabular}{cccc}
\hline
  UID &                                 Formula & Charge &    per cent \\
\hline
470 &        C$_{15}$H$_{9}$N$^{+}$ &      1 &          9.1 \\
 2482 &                      C$_{28}$ &      0 &          6.2 \\
  480 &  C$_{14}$H$_{11}$N$^{+}$ &      1 &          4.9 \\
 2758 &                      C$_{31}$ &      0 &          4.1 \\
 2598 &                      C$_{30}$ &      0 &          4.1 \\
 \vdots & \vdots & \vdots & \vdots  \\
\hline
\end{tabular}
\end{table}

\begin{table}
\centering
\caption{Same as Table  \ref{tab:emission_species_mrk273_Sy} but for  Mrk\,471.  The full table is available online.  }
\label{tab:emission_species_mrk471_Sy}
\begin{tabular}{cccc}
\hline
  UID &                                 Formula & Charge &    per cent \\
\hline
  3173 &              C$_{52}$H$_{18}$ &      0 &          9.0 \\
  836 &              C$_{54}$H$_{18}$ &      0 &          6.7 \\
 2427 &                           C$_{26}$ &      0 &          6.1 \\
  620 &              C$_{42}$H$_{16}$ &      0 &          5.7 \\
  233 &       C$_{23}$H$_{12}$N$^{+}$ &      1 &          5.0 \\
   \vdots & \vdots & \vdots & \vdots  \\
\hline
\end{tabular}
  \end{table}

\begin{table}
    \centering
    \caption{Same as Table  \ref{tab:emission_species_mrk273_Sy} but for  Mrk\,609.  The full table is available online.}
    \label{tab:emission_species_mrk609_Sy}
    \begin{tabular}{cccc}
\hline
  UID &                                 Formula & Charge &    \% \\
\hline
  3205 &   C$_{82}$H$_{24}^{-}$ &     -1 &          9.7 \\
  473 &              C$_{10}$H$_{9}$N &      0 &          7.7 \\
   76 &         C$_{47}$H$_{17}$ &      0 &          6.2 \\
  636 &   C$_{48}$H$_{18}^{+}$ &      1 &          5.3 \\
  128 &         C$_{36}$H$_{16}$ &      0 &          4.5 \\
  \vdots & \vdots & \vdots & \vdots  \\
\hline
\end{tabular}
\end{table}

\begin{table}
    \centering
     \caption{Same as Table  \ref{tab:emission_species_mrk273_Sy} but for  Mrk\,622. The full table is available online.}
    \label{tab:emission_species_mrk622_Sy}
   \begin{tabular}{cccc}
   \hline
  UID &                                 Formula & Charge &    \% \\
\hline
3173 &                C$_{52}$H$_{18}$ &      0 &         13.3 \\
  620 &                C$_{42}$H$_{16}$ &      0 &          7.8 \\
 2494 &                             C$_{28}$ &      0 &          3.6 \\
  233 &         C$_{23}$H$_{12}$N$^{+}$ &      1 &          3.5 \\
  724 &          C$_{45}$H$_{15}^{-}$ &     -1 &          3.3 \\
  \vdots & \vdots & \vdots & \vdots  \\
\hline
\end{tabular}
\end{table}

\begin{table}
    \centering
    \caption{Same as Table  \ref{tab:emission_species_mrk273_Sy} but for  Mrk\,883.  The full table is available online.}
    \label{tab:emission_species_mrk883_Sy}
   \begin{tabular}{cccc}
\hline
  UID &                                 Formula & Charge &    \% \\
\hline
   76 &          C$_{47}$H$_{17}$ &      0 &          7.5 \\
  473 &               C$_{10}$H$_{9}$N &      0 &          7.4 \\
 3205 &    C$_{82}$H$_{24}^{-}$ &     -1 &          6.0 \\
  625 &          C$_{40}$H$_{16}$ &      0 &          4.9 \\
  128 &          C$_{36}$H$_{16}$ &      0 &          4.5 \\
  \vdots & \vdots & \vdots & \vdots  \\

\hline
\end{tabular}
\end{table}

\begin{table}
\centering
    \caption{Same as Table  \ref{tab:emission_species_mrk273_Sy} but for  NGC\,660.  The full table is available online.}
    \label{tab:emission_species_ngc660_SB}
    \begin{tabular}{cccc}
\hline
  UID &                                 Formula & Charge &   \% \\
\hline
  470 &             C$_{15}$H$_{9}$N$^{+}$ &      1 &          5.0 \\
 3173 &              C$_{52}$H$_{18}$ &      0 &          4.5 \\
 2542 &                           C$_{29}$ &      0 &          4.3 \\
  128 &              C$_{36}$H$_{16}$ &      0 &          4.3 \\
  468 &             C$_{15}$H$_{9}$N$^{+}$ &      1 &          4.2 \\
  \vdots & \vdots & \vdots & \vdots  \\
\hline
\end{tabular}
\end{table}

\begin{table}
    \centering
     \caption{Same as Table  \ref{tab:emission_species_mrk273_Sy} but for NGC\,2622. The full table is available online.}
    \label{tab:emission_species_ngc2622_Sy}
    \begin{tabular}{cccc}
\hline
  UID &                            Formula & Charge &    \% \\
\hline
  620 &              C$_{42}$H$_{16}$ &      0 &         13.3 \\
  635 &              C$_{48}$H$_{18}$ &      0 &          6.8 \\
  233 &       C$_{23}$H$_{12}$N$^{+}$ &      1 &          6.8 \\
 3173 &              C$_{52}$H$_{18}$ &      0 &          6.4 \\
 2782 &                           C$_{32}$ &      0 &          4.7 \\
 \vdots & \vdots & \vdots & \vdots  \\
\hline
\end{tabular}
  \end{table}

\begin{table}
    \centering
        \caption{Same as Table  \ref{tab:emission_species_mrk273_Sy} but for  NGC\,2623. The full table is available online.}
    \label{tab:emssion_species_ngc2623_SB}
    \begin{tabular}{cccc}

\hline
  UID &                                 Formula & Charge &   \% \\
\hline
 2955 &                           C$_{33}$ &      0 &          5.7 \\
 3173 &              C$_{52}$H$_{18}$ &      0 &          5.4 \\
  473 &                   C$_{10}$H$_{9}$N &      0 &          4.9 \\
  468 &             C$_{15}$H$_{9}$N$^{+}$ &      1 &          4.6 \\
 2791 &                           C$_{32}$ &      0 &          4.5 \\
  \vdots & \vdots & \vdots & \vdots  \\
\hline
\end{tabular}
\end{table}

\begin{table}
    \centering
    \caption{Same as Table  \ref{tab:emission_species_mrk273_Sy} but for  NGC\,4676. The full table is available online.}
\label{tab:emission_species_ngc4676_SB}
   \begin{tabular}{cccc}
\hline
  UID &                                 Formula & Charge &   \% \\
\hline
  3173 &               C$_{52}$H$_{18}$ &      0 &          8.6 \\
  473 &                    C$_{10}$H$_{9}$N &      0 &          6.8 \\
  636 &         C$_{48}$H$_{18}^{+}$ &      1 &          5.6 \\
  128 &               C$_{36}$H$_{16}$ &      0 &          4.9 \\
  468 &              C$_{15}$H$_{9}$N$^{+}$ &      1 &          4.5 \\
   \vdots & \vdots & \vdots & \vdots  \\
\hline
\end{tabular}
\end{table}
\begin{table}
    \centering
    \caption{Same as Table  \ref{tab:emission_species_mrk273_Sy} but for  NGC\,4922. The full table is available online.}
\label{tab:emission_species_ngc4922_Sy}
 \centering
\begin{tabular}{cccc}
\hline
  UID &                            Formula & Charge &        \% \\
\hline
3205 &   C$_{82}$H$_{24}^{-}$ &     -1 &          8.1 \\
 3173 &         C$_{52}$H$_{18}$ &      0 &          6.9 \\
  836 &         C$_{54}$H$_{18}$ &      0 &          5.8 \\
  636 &   C$_{48}$H$_{18}^{+}$ &      1 &          4.2 \\
  473 &              C$_{10}$H$_{9}$N &      0 &          4.1 \\
   \vdots & \vdots & \vdots & \vdots  \\
 \hline
\end{tabular}
 \end{table}

\begin{table}
    \centering
    \caption{Same as Table  \ref{tab:emission_species_mrk273_Sy} but for   NGC\,5256. The full table is available online.}
\label{tab:emission_species_ngc5256_Sy}
 \centering
\begin{tabular}{cccc}
\hline
  UID &                            Formula & Charge &        \% \\
\hline
473 &                    C$_{10}$H$_{9}$N &      0 &          6.0 \\
  822 &         C$_{34}$H$_{20}^{-}$ &     -1 &          4.9 \\
  636 &         C$_{48}$H$_{18}^{+}$ &      1 &          4.9 \\
 3205 &         C$_{82}$H$_{24}^{-}$ &     -1 &          4.8 \\
 3173 &               C$_{52}$H$_{18}$ &      0 &          4.4 \\
  \vdots & \vdots & \vdots & \vdots  \\
\hline
\end{tabular}
 \end{table}

\begin{table}    
\centering
         \caption{Same as Table  \ref{tab:emission_species_mrk273_Sy} but NGC\,5347. The full table is available online.} 
    \label{tab:emission_species_ngc347_Sy}
    \begin{tabular}{cccc}
\hline
  UID &                                 Formula & Charge &    \% \\
  \hline
 3173 &               C$_{52}$H$_{18}$ &      0 &          6.3 \\
  480 &        C$_{14}$H$_{11}$N$^{+}$ &      1 &          6.0 \\
 2758 &                            C$_{31}$ &      0 &          5.8 \\
  473 &                    C$_{10}$H$_{9}$N &      0 &          5.4 \\
 2955 &                            C$_{33}$ &      0 &          4.3 \\
  \vdots & \vdots & \vdots & \vdots  \\
\hline
\end{tabular}
\end{table}

\section{Peeters' classification procedure}
\label{ap:peeters}
\let\cleardoublepage\clearpage
The PAH band profile variations were studied in several astrophysical objects by \citet{peeters2002rich}, which separated those bands intro three different classes (A, B and C), according to their  central wavelengths. The 6 -- 9~$\mu$m spectral region, in special, is composed by three PAH main features: a band at 6.2~$\mu$m, a complex of overlapping bands at 7.7~$\mu$m with  two components at 7.6 and 7.8~$\mu$m, and a band at 8.6~$\mu$m \citep{Ricca18}. In the case of the first band, the profile A peaks at shorter wavelengths in comparison to B and C profiles. For the 7.7~$\mu$m complex, the classes A and B differ in the relative strength of the 7.6 and 7.8~$\mu$m features (F$_{7.6}$/F$_{7.8}$ flux ratio), which seem to be shifted to a peak position of 8.2~$\mu$m for class C objects \citep{tielens2008interstellar}. Finally, the 8.6~$\mu$m band also peaks at shorter wavelengths for A profiles.

\begin{figure*}
    \centering
    \includegraphics[scale=0.38]{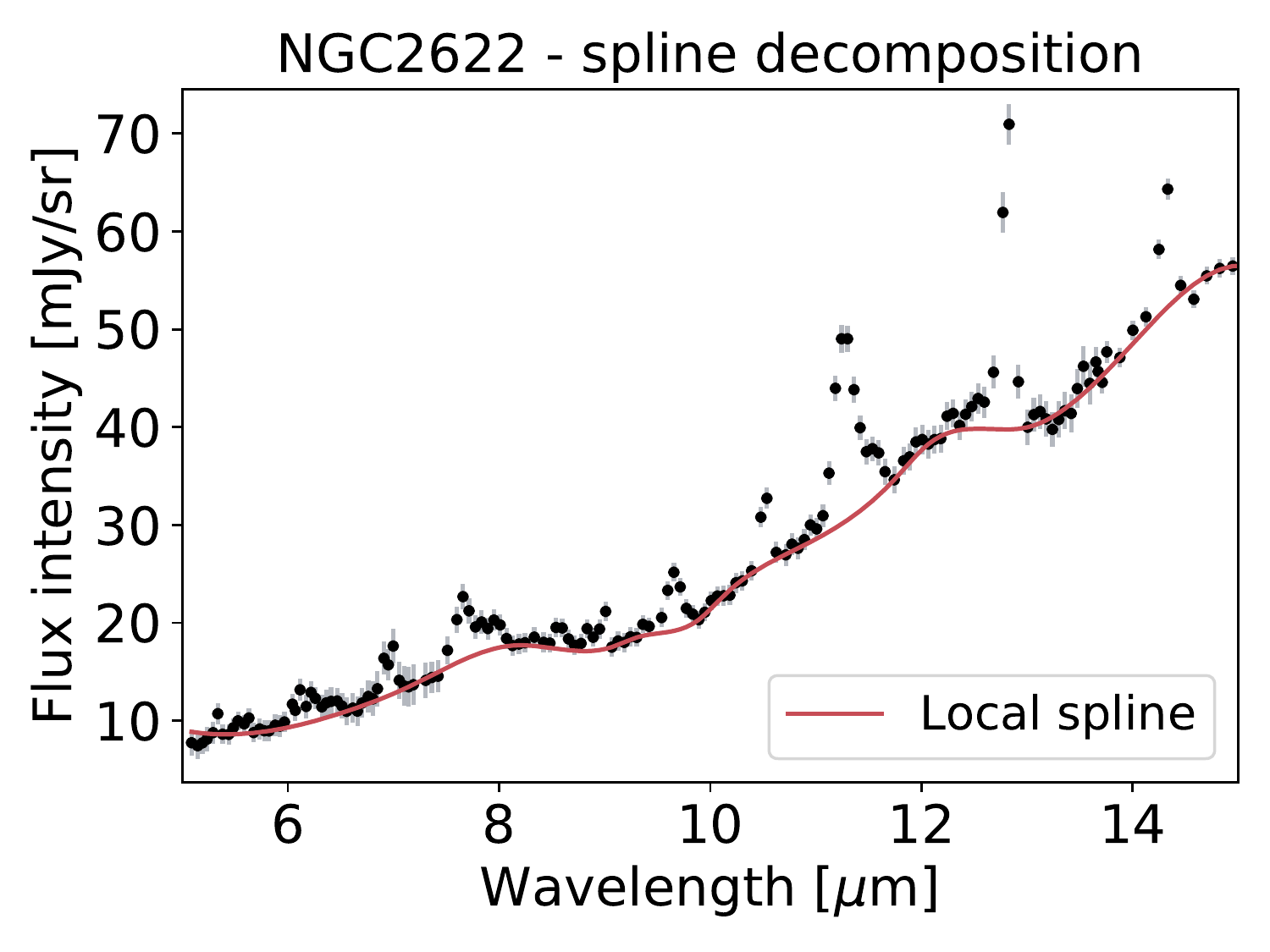}
    \includegraphics[scale=0.38]{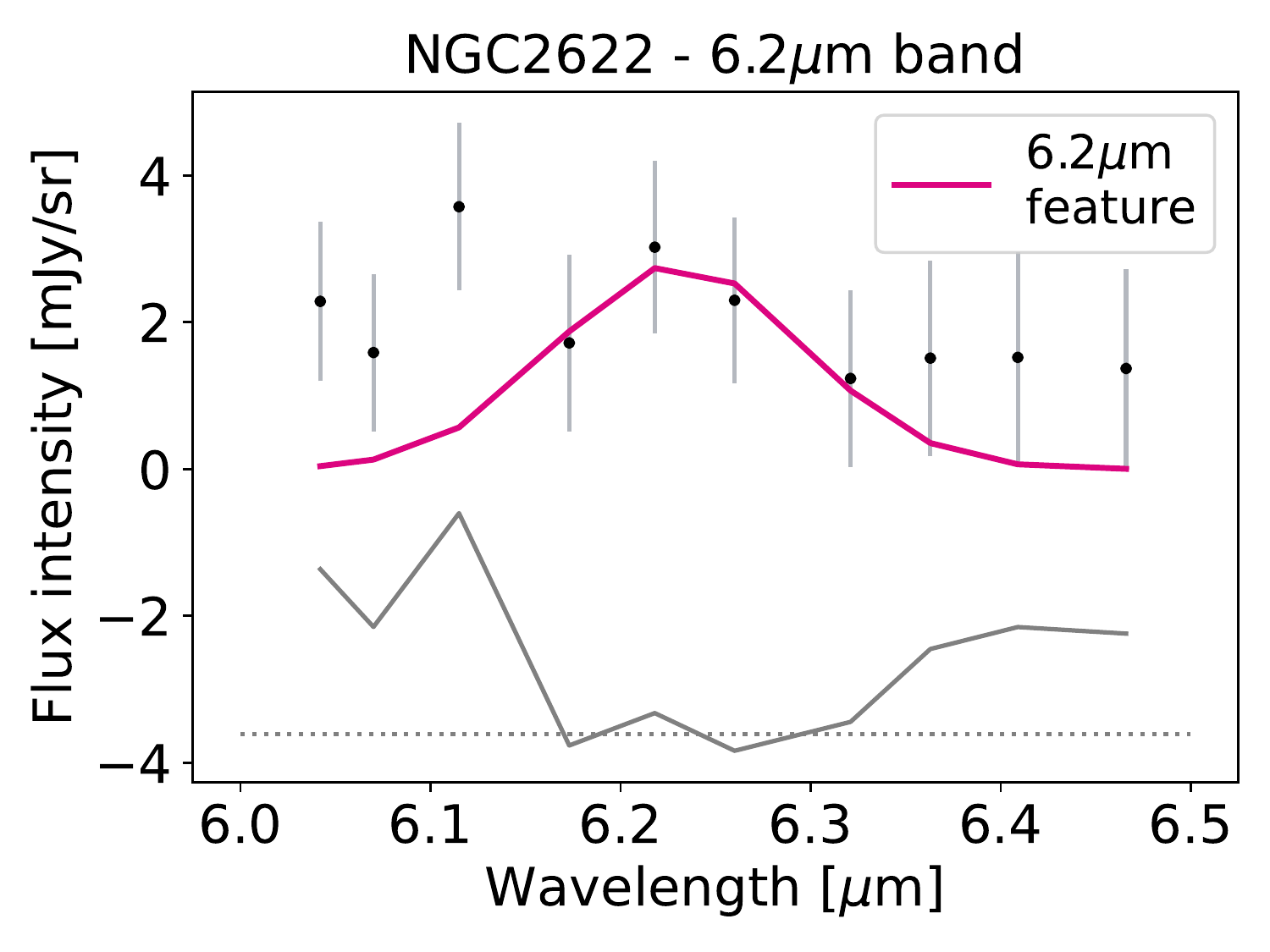}
    \includegraphics[scale=0.38]{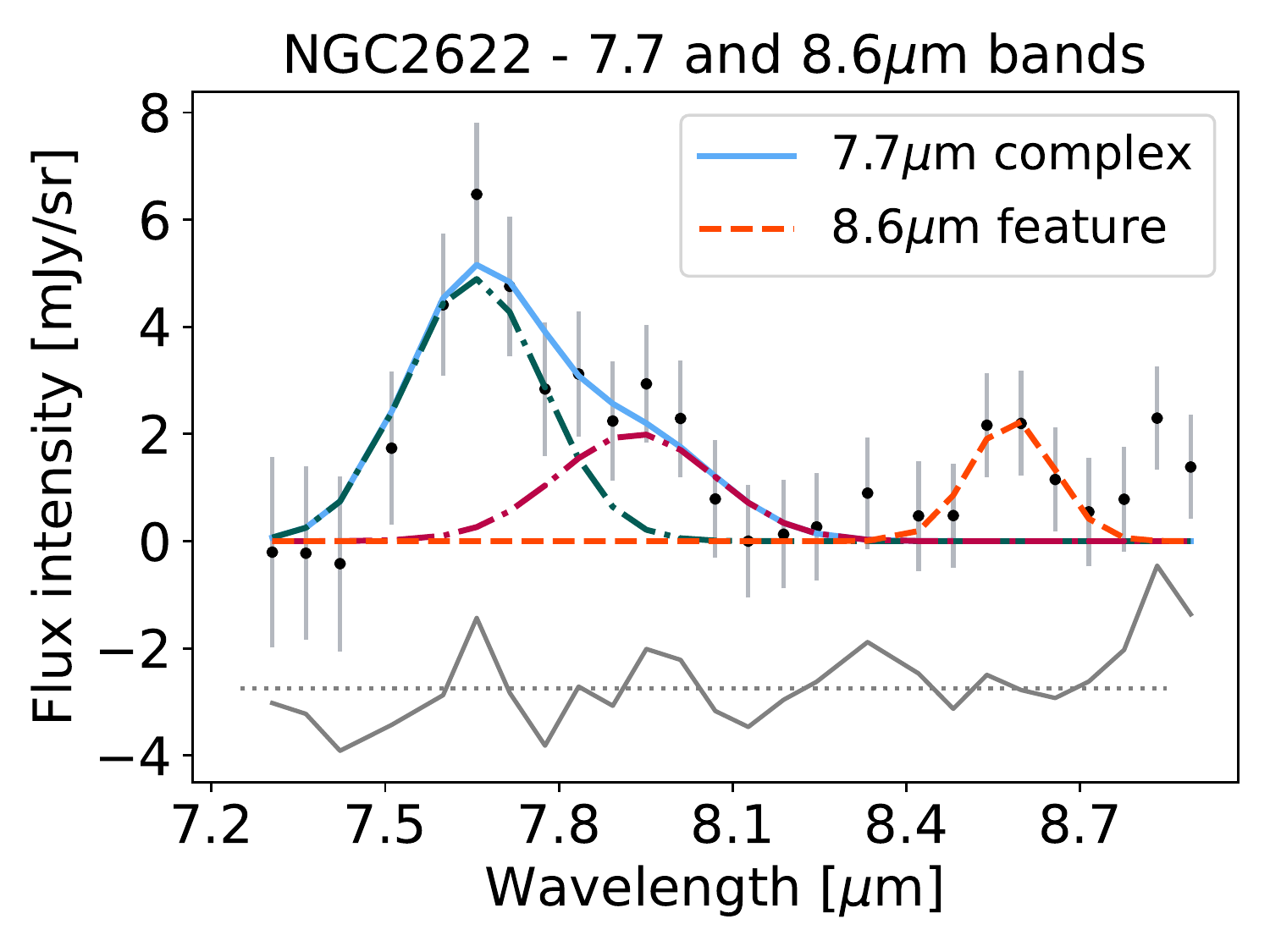}
    
    \includegraphics[scale=0.38]{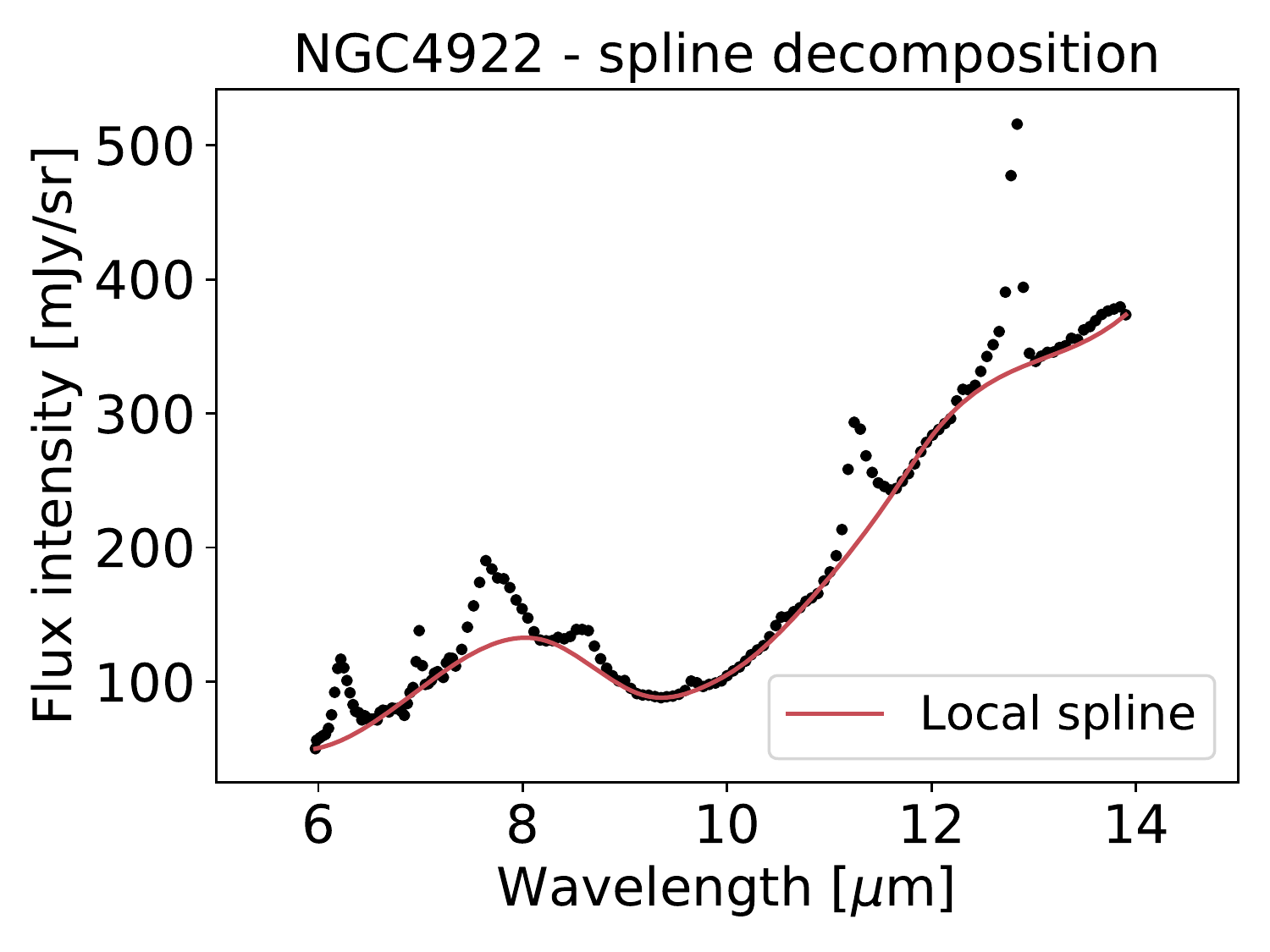}
    \includegraphics[scale=0.38]{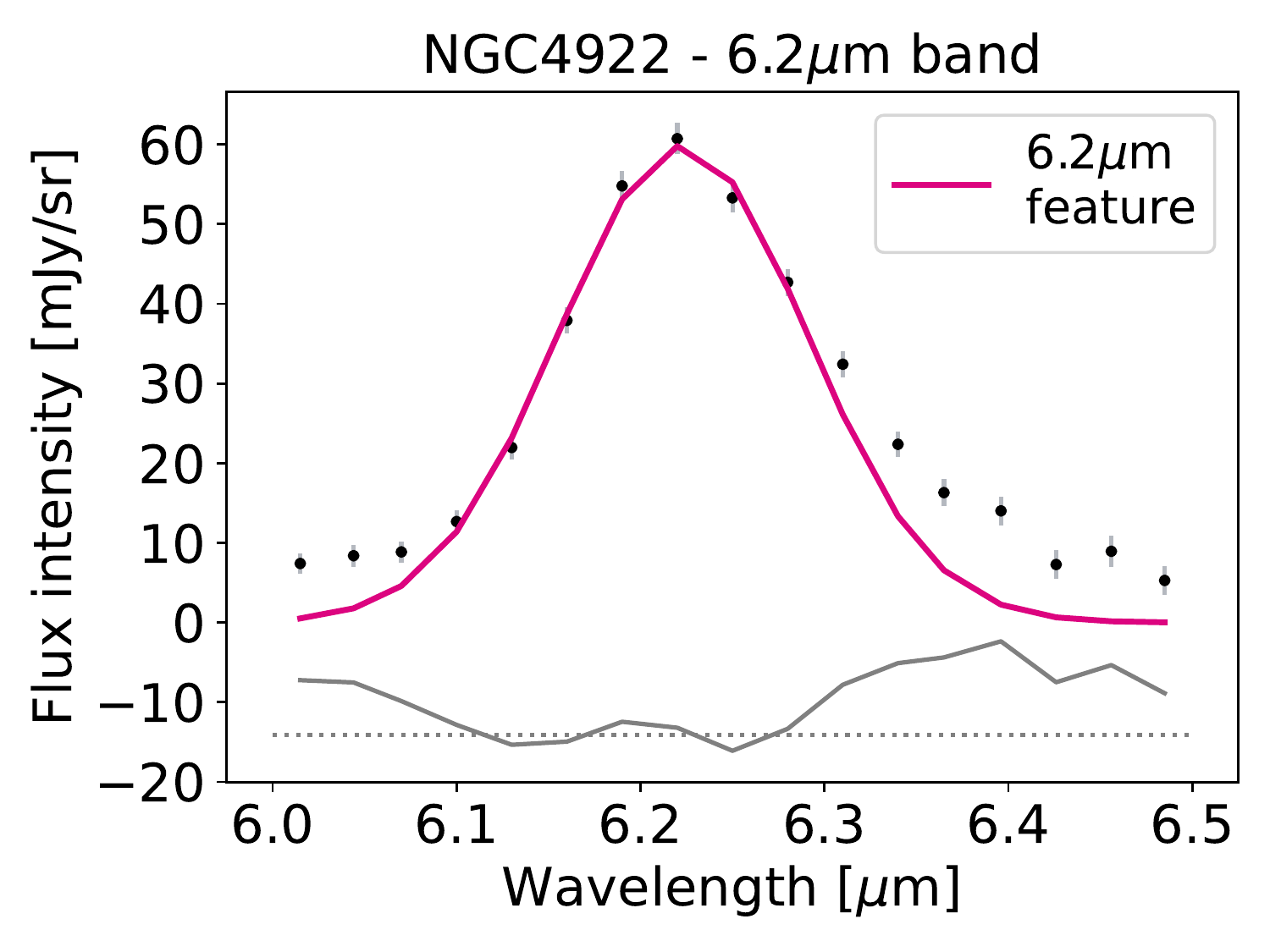}
    \includegraphics[scale=0.38]{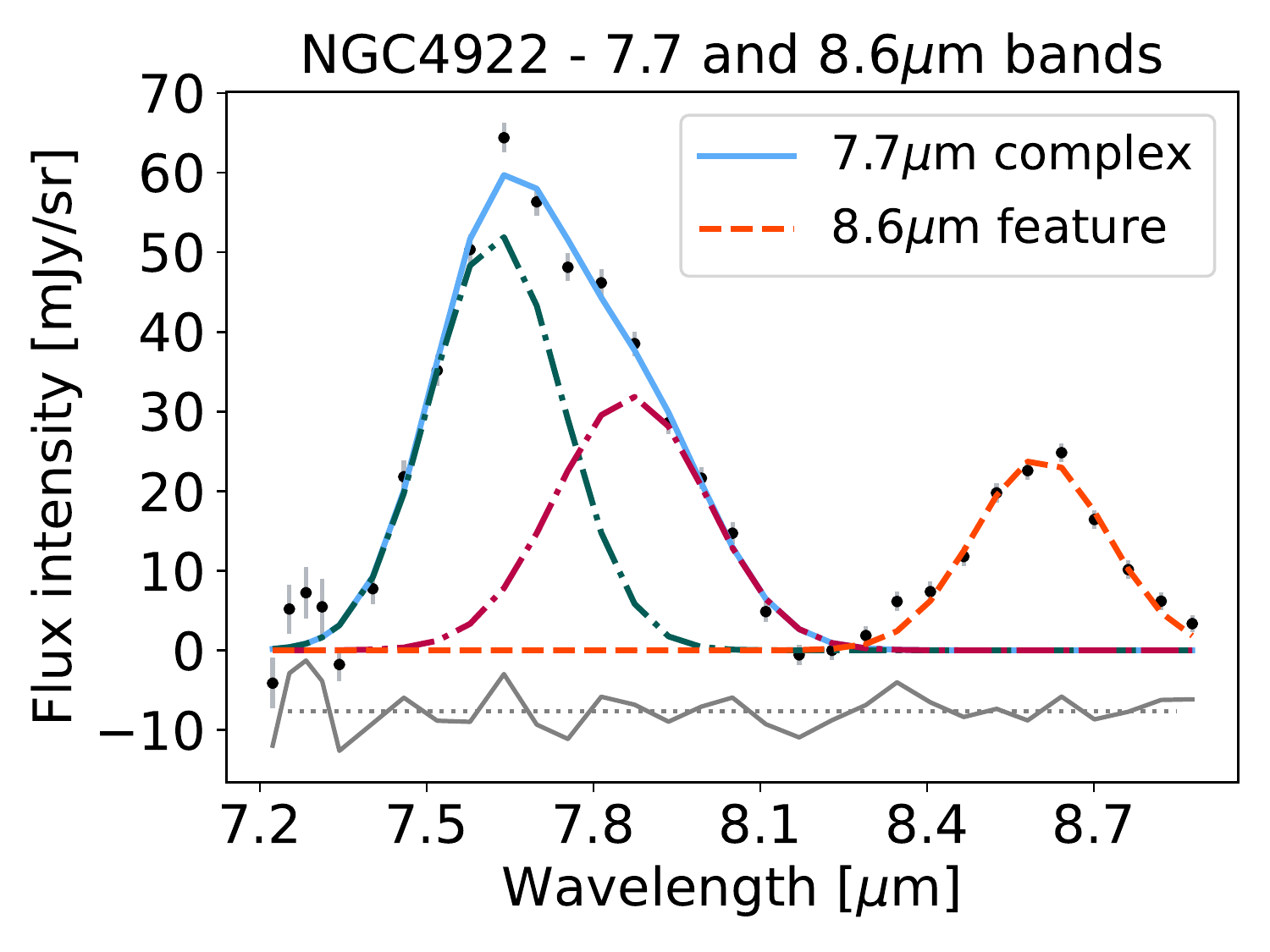}
    
    \includegraphics[scale=0.38]{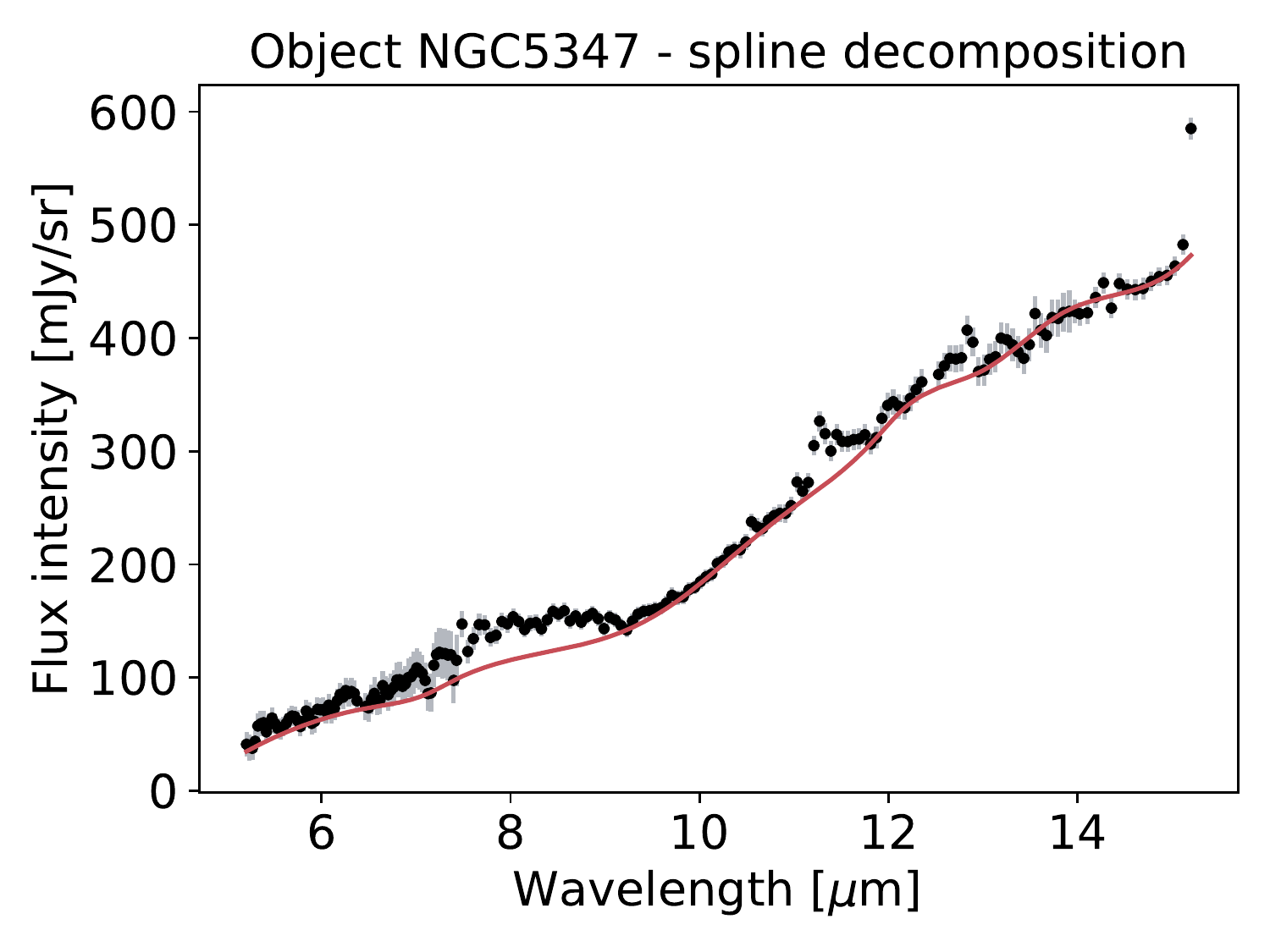}
    \includegraphics[scale=0.38]{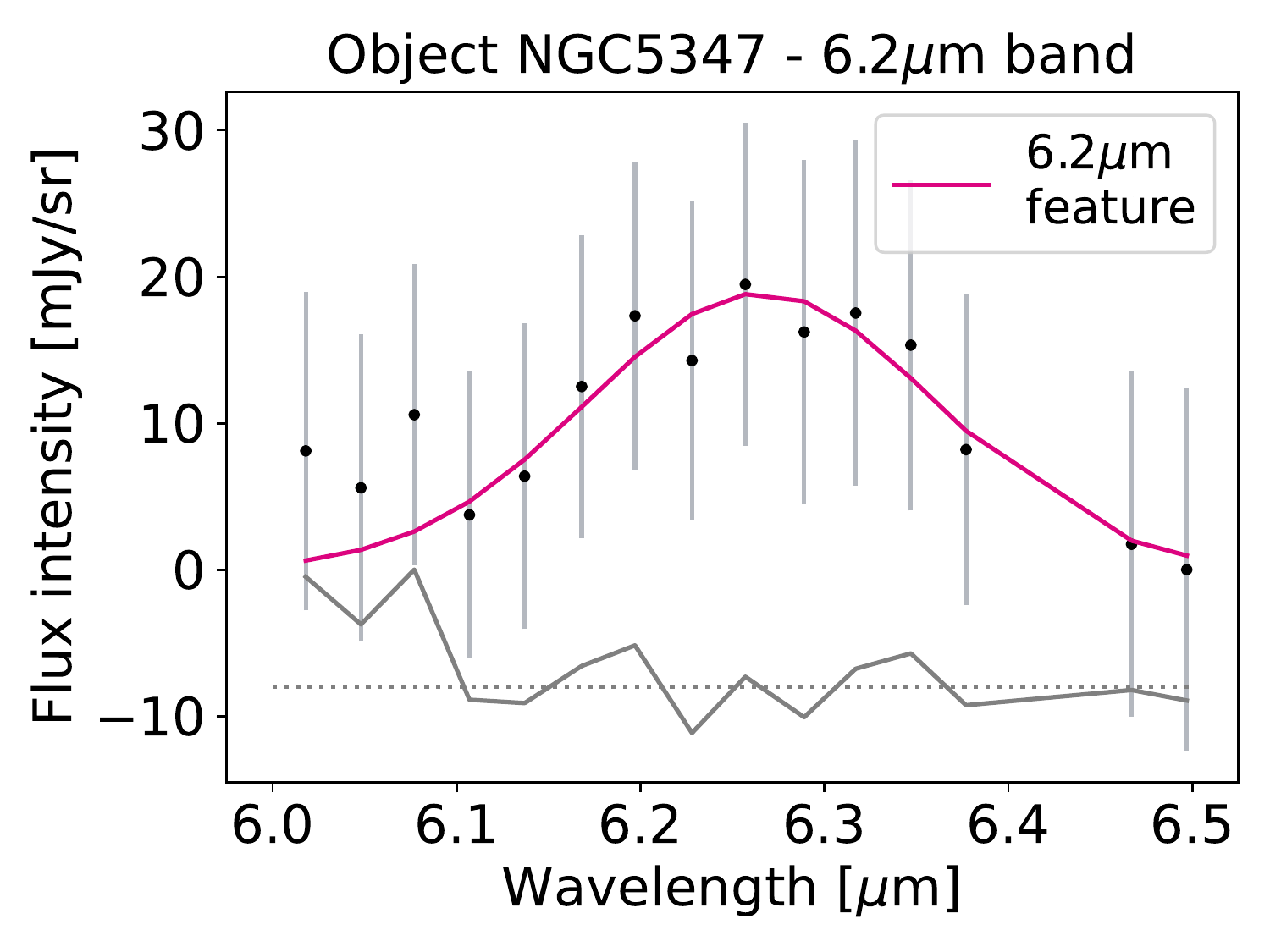}
    \includegraphics[scale=0.38]{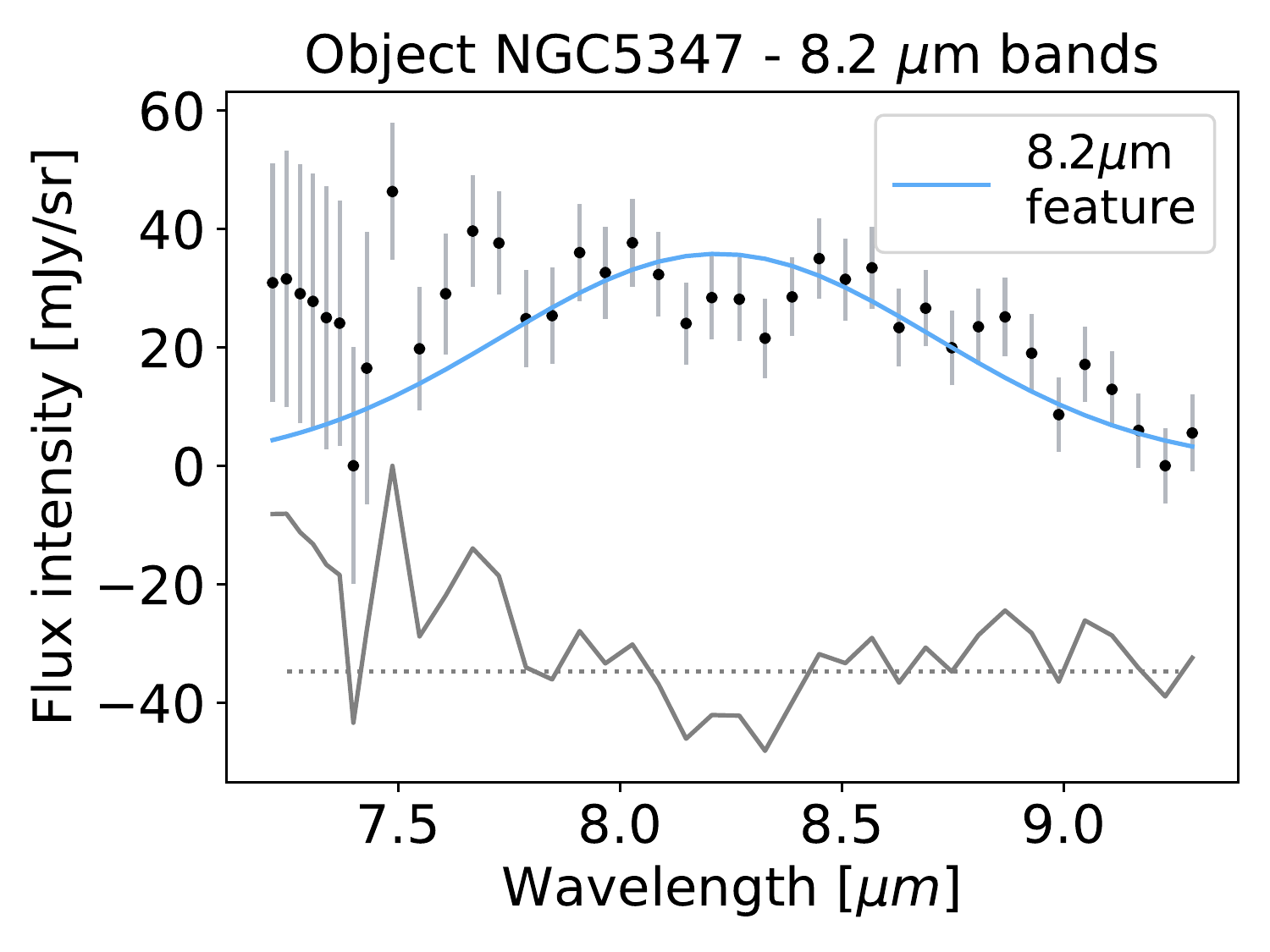}

    \caption{Local spline decomposition and fit results of the 6.2, 7.7 and 8.6 bands for the galaxies of NGC~2622, NGC~4922 and  NGC~5347, from top to down. The data points and error bars are in black and grey, respectively. }
    \label{fig:galaxies-python-fits}
\end{figure*}

The methodology applied to classify the NGC~2622, NGC~4922 and NGC~5347 galaxies is the same used in the other objects of our sample, previously performed by \citet{canelo2018variations},  \citet{thesis-canelo} and \citet{Canelo2021}. The {\sc PAHFIT} decomposition does not allow the band central wavelengths to vary and can not be used to this analysis. Therefore, we subtracted the underlying continuum with a spline decomposition \citep[e.g.][]{galliano2008variations,peeters2017pah}, with a set of anchor points at roughly 5.4, 5.8, 6.6, 7.2, 8.2, 9.0, 9.3, 9.9, 10.2, 10.9, 11.7, 12.1, 13.1, 13.9, 14.7 and 15.0~$\mu$m. In sequence, the 6.2, 7.7 and 8.6~$\mu$m bands were independently fitted with a Gaussian profile \citep[based on the procedure of][for instance]{peeters2017pah}, with a \textsc{python}-based script constructed to estimate their central wavelength ($\lambda_c$), amplitude and FWHM through the optimisation algorithms from the submodule \textit{scipy.optmize.curve\_fit}. The uncertainties of these parameters were also derived by this tool with least-squares minimisation from the flux uncertainties provided by the ATLAS. The initial guesses for the parameters were selected from \citet{smith2007mid}. The continuum and bands fitting are shown in Fig.~\ref{fig:galaxies-python-fits} and Table
~\ref{tab:fits-gauss}.  In the specific case of the 7.7~$\mu$m complex, we fixed the FWHM values according to \citet{peeters2002rich} in 0.28 to 7.6~$\mu$m and 0.32 to 7.8~$\mu$m profiles, respectively,  to avoid the blending of the features.  

The particular case of NGC~5347 needed a slightly different approach. The PAH emissions are weaker in this galaxy and the 7~--~10~$\mu$m region resembles an unique bump feature, as expected in class C objects.  As a matter of fact, this source was already pointed out as Compton-thick AGN, with a dusty material obscuring the central engine \citep{Kammoun2019}. In order to proper analyse NGC~5347, the anchor point at 8~$\mu$m was not included in the spline decomposition and the 7.7 and 8.6~$\mu$m bands were fitted with just one broad Gaussian profile with peak position at roughly 8.22~$\mu$m \citep[see, for example,][]{peeters2002rich}. In such objects, the 7.7~$\mu$m corresponds to a class C object and there is no emission at 8.6~$\mu$m.

With the results of the Gaussian fits, it is possible to separate the sources into the Peeters' classes.  The 6.2~$\mu$m band is considered as class A if $\lambda_c$~<~6.23~$\mu$m, class B if 6.23~$\mu$m < $\lambda$ < 6.29~$\mu$m, and class C if  $\lambda_c$~>~6.29. The 7.7~$\mu$m complex is considered as  class A if F$_{7.6}$/F$_{7.8}$~$\geq$~1, class B if F$_{7.6}$/F$_{7.8}$~<~1, and class C if $\lambda_c \sim$~8.22~$\mu$m (the 7.6 and 7.8~$\mu$m features are not present). The 8.6~$\mu$m band does not have a class C profile, and the A and B classifications depend on whether $\lambda_c$~<~8.60~$\mu$m or $\lambda_c$~>~8.60~$\mu$m, respectively. The classification of NGC~2622, NGC~4922 and NGC~5347 are displayed in Table~\ref{tab:carla-classes}. The F$_{7.6}$/F$_{7.8}$ ratios of NGC~2622 and NGC~4922 are 2.187~$\pm$~0.513 and 1.444~$\pm$~0.134, respectively. 

\begin{table}
\centering
\caption{Best-fit results for the 6.2, 7.7 and 8.6~$\mu$m bands. A is the amplitude in mJy/sr, $\lambda_c$ is the central wavelength in $\mu$m and FWHM is the full width at half maximum.}
\label{tab:fits-gauss}
{\footnotesize
\begin{tabular}{ccccccc}
\hline
Source & $\lambda_c$ & Err & A & Err & FWHM & Err \\
\hline
NGC 2622 & 6.231 & 0.011 & 0.455 & 0.078 & 0.153 & 0.032 \\
 & 7.653 & 0.024 & 1.460 & 0.171 & 0.280 & --- \\
 & 7.932 & 0.051 & 0.684 & 0.166 & 0.320 & --- \\
 & 8.582 & 0.014 & 0.413 & 0.069 & 0.170 & 0.033 \\
NGC 4922 & 6.223 & 0.002 & 10.130 & 0.248 &  0.159 & 0.005 \\
 & 7.625 & 0.009 & 15.574 & 0.845 & 0.280 & --- \\
 & 7.867 & 0.014 & 10.859 & 0.844 & 0.320 & --- \\
 & 8.604 & 0.005 & 7.271 & 0.312 & 0.283 & 0.015 \\
NGC 5347 &  6.266 & 0.007 & 4.488 & 0.290 & 0.223 & 0.018 \\
 & 8.221 & 0.048 & 43.786 & 3.868 & 1.150 & 0.122 \\
\hline 
\end{tabular}
}
\end{table}

\bsp
\label{lastpage}
\end{document}